\documentclass[journal,comsoc]{IEEEtran}

\usepackage{cite}
\usepackage[caption=false,font=footnotesize,labelfont=sf,textfont=sf]{subfig}
\usepackage{balance}
\usepackage{tabularx}
\usepackage{booktabs}
\usepackage{amsmath,amssymb,amsfonts}
\usepackage{mathtools} 
\usepackage{graphicx}
\usepackage{xcolor}
\usepackage{tikz}
\usepackage{float}
\usepackage{adjustbox}
\usepackage{diagbox} 
\usepackage[nolist,nohyperlinks]{acronym}
\usepackage{algorithm}
\usepackage{pifont}
\usepackage{algpseudocode}
\usepackage{setspace}
\let\Algorithm\algorithm
\renewcommand\algorithm[1][]{\Algorithm[#1]\setstretch{1.1}}

\newcommand{\av}{\mathbf{a}}
\newcommand{\bv}{\mathbf{b}}

\newcommand{\bh}{\mathbf{b}^\herm}
\newcommand{\ap}{\mathbf{\dot{a}}}
\newcommand{\bp}{\mathbf{\dot{b}}}
\newcommand{\bph}{\mathbf{\dot{b}}^\herm}
\newcommand{\bpp}{\mathbf{\ddot{b}}}
\newcommand{\bpph}{\mathbf{\ddot{b}}^\herm}
\newcommand{\thetat}{\theta_{\mathrm{T}}}
\newcommand{\thetar}{\theta_{\mathrm{R}}}
\newcommand{\thetarn}{\theta_{\mathrm{R},n}}
\newcommand{\thetaln}{\theta_{\mathrm{L},n}}
\newcommand{\thetan}{\vartheta_{n}}

\newcommand{\Thetatwo}{\boldsymbol{\Theta}_2}
\newcommand{\Thetanone}{\boldsymbol{\Theta}_{n,1}}
\newcommand{\Thetantwo}{\boldsymbol{\Theta}_{n,2}}
\newcommand{\Na}{N_{\mathrm{a}}}
\newcommand{\Nt}{N_{\mathrm{T}}}
\newcommand{\Nr}{N_{\mathrm{R}}}
\newcommand{\Nbs}{N_{\mathrm{BS}}}
\newcommand{\h}{\mathbf{h}}
\newcommand{\rv}{\mathbf{r}}
\newcommand{\hchan}{\mathbf{H}}
\newcommand{\hh}{\mathbf{h}^\herm}
\newcommand{\y}{\mathbf{y}}
\newcommand{\vv}{\mathbf{v}}
\newcommand{\yh}{\mathbf{y}^\herm}
\newcommand{\fd}{f_\mathrm{D}}

\newcommand{\fdn}{f_{\mathrm{D},n}}
\newcommand{\Ts}{T_\mathrm{s}}
\newcommand{\Df}{\Delta f}

\newcommand{\xbs}{x^{(n)}_{\mathrm{bs}}}

\newcommand{\ybs}{y^{(n)}_{\mathrm{bs}}}

\newcommand{\EX}[1] {{\mathbb{E}}\left\{{#1}\right\}}
\newcommand{\p}{\mathbf{p}}
\newcommand{\ph}{\widehat{\mathbf{p}}}
\newcommand{\I}{\boldsymbol{\mathcal{I}}}
\newcommand{\Ie}{\boldsymbol{\mathcal{I}}_\mathrm{e}}
\newcommand{\Jm}{\mathbf{J}_\Xi}
\newcommand{\Jinvm}{\mathbf{J}_{\Xi^{-1}}}
\newcommand{\Jn}{\mathbf{J}_\mathrm{n}}

\newcommand{\SNR}{\mathrm{SNR}}
\newcommand{\CRB}{\mathrm{CRLB}}
\newcommand{\herm}{\mathsf{H}}
\newcommand{\transp}{\mathsf{T}}

\newcounter{MYtempeqncnt}

\DeclareMathOperator{\diag}{diag}
\DeclareMathOperator{\Tr}{Tr}

\begin{acronym}
\acro{AWGN}{additive white Gaussian noise}
\acro{AoA}{angle of arrival}
\acro{BF}{beamformer}
\acro{BS}{base station}
\acro{CP}{cyclic prefix}
\acro{CPU}{central processing unit}
\acro{CRLB}{Cram\'{e}r-Rao lower bound}
\acro{DoA}{direction of arrival}
\acro{DoD}{direction of departure}
\acro{EIRP}{effective isotropic radiated power}
\acro{ELP}{equivalent low-pass}
\acro{EFIM}{equivalent Fisher information matrix}
\acro{FFT}{fast Fourier transform}
\acro{FIM}{Fisher information matrix}
\acro{ICI}{inter-carrier interference}
\acro{IFFT}{inverse fast Fourier transform}
\acro{i.i.d.}{independent, identically distributed}
\acro{ISAC}{integrated sensing and communication}
\acro{ISI}{inter-symbol interference}
\acro{JSC}{joint sensing and communication}
\acro{LOS}{line-of-sight}
\acro{LTE}{long term evolution}
\acro{LFM}{linear frequency modulated}
\acro{ML}{maximum likelihood}
\acro{MCL}{maximum coupling loss}
\acro{MPL}{maximum path loss}
\acro{MIL}{maximum isotropic loss}
\acro{MC}{Monte Carlo}
\acro{MIMO}{multiple-input multiple-output}
\acro{MUSIC}{MUltiple SIgnal Classification}
\acro{MDL}{minimum description length}
\acro{NR}{new radio}
\acro{OFDM}{orthogonal frequency division multiplexing}
\acro{OSPA}{optimal sub-pattern assignment}
\acro{PEB}{position error bound}
\acro{PSD}{power spectral density}
\acro{QPSK}{quadrature phase shift keying}
\acro{RCS}{radar cross-section}
\acro{Rx}{receiver}
\acro{RF}{radio frequency}
\acro{RMSE}{root mean square error}
\acro{r.v.}{random variable}
\acro{SSIR}{signal-to-self interference ratio}
\acro{SI}{self interference}
\acro{SNR}{signal-to-noise ratio}
\acro{SU}{single-user}
\acro{SCM}{sample covariance matrix}
\acro{Tx}{transmitter}
\acro{UE}{user equipment}
\acro{ULA}{uniform linear array}
\acro{UMTS}{universal mobile telecommunication system}
\acro{VEB}{velocity error bound}
\end{acronym}

\begin{document}

\title{Position and Velocity Estimation Accuracy in MIMO-OFDM ISAC Networks:\\ A Fisher Information Analysis}

\author{Lorenzo Pucci,~\IEEEmembership{Member,~IEEE}, Luca Arcangeloni,~\IEEEmembership{Member,~IEEE}, and Andrea~Giorgetti,~\IEEEmembership{Senior~Member,~IEEE}
\thanks{This work has been submitted to the IEEE for possible publication. Copyright may be transferred without notice, after which this version may no longer be accessible.}
\thanks{This work has been presented, in part, at the IEEE Int. Work. on Signal Proc. Adv. in Wireless Comm.(SPAWC), Lucca, Italy, Sep. 2024 \cite{PucGio:C24}.}\\
\thanks{This work was supported 
by the European Union under the Italian National Recovery and Resilience Plan (NRRP) of NextGenerationEU, partnership on ``Telecommunications of the Future'' (PE00000001 - program ``RESTART'').}
\thanks{The authors are with the Department of Electrical, Electronic, and Information Engineering ``Guglielmo Marconi'' and CNIT/WiLab, University of Bologna, Italy (e-mail: \{lorenzo.pucci3, luca.arcangeloni2, andrea.giorgetti\}@unibo.it). }
}
\maketitle

\begin{abstract}
This paper presents a theoretical framework to derive information-theoretic bounds on the estimation accuracy of target position and velocity in \ac{OFDM}-based \ac{ISAC} networks composed of multiple cooperative and distributed \ac{MIMO} \acp{BS}. Leveraging Fisher information analysis, we derive closed-form expressions for the \acp{CRLB} in both monostatic and bistatic configurations. The framework is then extended to cooperative settings, including networks with multiple coordinated monostatic sensors and multistatic configurations, enabling joint estimation of target position and velocity. We systematically examine how estimation accuracy depends on key system parameters such as the number of \acp{BS}, bandwidth, antenna configuration, and network geometry. Numerical results highlight the performance gains enabled by cooperative sensing and provide insights to guide the design of future \ac{ISAC} systems. 
\end{abstract}
\acresetall


\section{Introduction}
The upcoming sixth-generation (6G) mobile networks are expected to significantly increase communication capacity and reduce latency, while natively supporting sensing applications. \Ac{ISAC} represents a paradigm shift by enabling the same radio infrastructure to simultaneously transmit data and sense the physical environment.
This convergence underpins a wide range of future applications, including autonomous transportation, smart cities, and human-machine interfaces, all of which demand real-time tracking, situational awareness, and centimeter-level positioning accuracy \cite{Bartoletti:M24,Decarli:J24}.
By sharing spectral and hardware resources between communication and sensing, \ac{ISAC} also improves spectral efficiency, lowers system costs, and enables tighter integration between networks and the physical world. This is crucial for meeting the demands of a hyper-connected society \cite{LiuHuaLi:J22, MatFavPucXuPaoGio:M25}.

Among candidate waveforms, \ac{OFDM} stands out for \ac{ISAC} due to its flexibility across time, frequency, and spatial domains. Widely adopted in fourth-generation (4G), fifth-generation (5G), and Wi-Fi, \ac{MIMO}-\ac{OFDM} enables both reliable communication and high-resolution sensing \cite{HuaZhaetal:J25}. Consequently, recent studies have evaluated the sensing performance of \ac{OFDM}-based \ac{ISAC} systems using metrics such as \ac{RMSE} and \ac{CRLB} \cite{Braun, TagMizetal:J23, PucMatPaoGio:C22}.

Despite these efforts, most prior \ac{CRLB}-based studies focus on simplified scenarios, analyzing individual parameters (e.g., delay, angle, Doppler) in monostatic or bistatic setups. However, a unified framework for joint position and velocity estimation in \ac{ISAC} networks is still lacking.
This gap is increasingly relevant as future mobile networks are expected to leverage spatially distributed infrastructure, enabling cooperative sensing in both monostatic and bistatic modes to enhance coverage, robustness, and situational awareness \cite{StrAleAma:L24, KaiHanMas:C25, KaiMas:C24,FavMatPucPaoXuGio:J25,DehPucJunGioPaoCai:J24}.

Understanding the fundamental performance limits of cooperative \ac{ISAC} is essential for benchmarking algorithms and guiding system-level design choices, such as resource allocation, \ac{BS} placement, and cooperation strategies \cite{ManTagTebMizetal:J25}. As \ac{ISAC} becomes a practical foundation of wireless networks, quantitatively characterizing these limits is critical to avoid inefficiencies and performance bottlenecks.

\subsection{Existing Works}
Several studies have investigated the performance of \ac{ISAC} systems using the \ac{CRLB}, primarily focusing on single monostatic or bistatic configurations. In \cite{liu2021cramer,ZabPaoXuGio:C22} and \cite{ZabPaoXuGio:C24}, narrowband monostatic \ac{ISAC} systems, leveraging \ac{AoA} estimation only, are analyzed, where the \ac{CRLB} serves as a metric to quantify the trade-off between communication and sensing accuracy. Similarly, \cite{Giovannetti2024PEB} derives the \ac{CRLB} for position estimation in a vehicular scenario, also under a monostatic \ac{ISAC} setup.

In \cite{FanNguJun:C25}, the authors propose a low-complexity beamforming scheme for monostatic multi-user \ac{ISAC} systems that jointly optimizes the communication sum-rate and sensing accuracy, as quantified by the \ac{CRLB}. A single \ac{BS} equipped with colocated transmit and receive antenna arrays is considered, simultaneously serving multiple single-antenna users and sensing a point-like target.

Bistatic configurations have also been examined. In \cite{BistaticBoundsLFM}, the impact of system geometry and transmitted waveforms on the ambiguity function is analyzed in the context of a bistatic radar channel. The \ac{CRLB} is employed to assess estimation accuracy and to identify optimal bistatic \ac{Tx}–\ac{Rx} pairs for target tracking in a multistatic radar system. The analysis assumes transmission of a burst of \ac{LFM} pulses.

The work in \cite{XuXieXu:C23} investigates the fundamental performance limits of collaborative sensing in perceptive mobile networks by deriving the \ac{CRLB}  and analyzing the impact of system parameters such as the number of \acp{BS}, antennas, and target position relative to the \acp{BS}. The authors also propose a \ac{BS} selection algorithm to balance localization accuracy and complexity. However, the provided analysis does not consider \ac{OFDM} signaling and is limited to targets located along a straight road.
In \cite{YaoZhaCai:C23}, the authors derive the \ac{CRLB} for single-target localization in a cloud radio access network system using \ac{OFDM} signals, assuming single-antenna devices. A closed-form  \ac{CRLB} is obtained from the \ac{FIM}, capturing the impact of system parameters like subcarrier number, transmit power, and \ac{Rx} geometry on localization accuracy. 
In \cite{SakGueBou:C21}, the authors investigate the problem of localizing a point scatterer using a \ac{MIMO}-\ac{OFDM} distributed system, where the communication waveform is reused for sensing. The \ac{CRLB} is derived for this multistatic system, and the theoretical performance is validated through both simulations and experiments.

With regard to velocity estimation accuracy, several studies have addressed related challenges from different perspectives. In \cite{WanMuLiu:J25}, the authors investigate \ac{ML} target velocity estimation and its application to predictive beamforming. However, this work does not analyze fundamental performance bounds or examine how system or geometric parameters affect estimation accuracy. In contrast, \cite{CatNicDav:L24} considers extremely large aperture arrays in a monostatic configuration and derives analytical performance bounds for both the radial and transverse components of target velocity. Additionally, \cite{HeBluHai:J10} derives the \ac{CRLB} for velocity estimation in \ac{MIMO} radar systems with widely spaced antennas, and proposes optimal antenna placement strategies to minimize estimation error.

\subsection{Our Contribution}
To the best of the authors' knowledge, the fundamental performance limits of joint position and velocity estimation in \ac{ISAC} networks, whether comprising multiple cooperative monostatic sensors, configured as multistatic systems, or operating under mixed configurations, have not been comprehensively characterized. In particular, a unified analysis that quantifies these limits under distributed cooperation among sensing nodes remains largely unexplored. Moreover, the influence of network geometry and resource allocation on estimation accuracy has yet to be rigorously analyzed.

This paper addresses these gaps by investigating the fundamental limits of target localization and velocity estimation in \ac{OFDM}-based \ac{ISAC} networks composed of multiple \ac{MIMO} \acp{BS}. These \acp{BS} cooperatively sense the environment to estimate target position and velocity, both in magnitude and direction, using a portion of the available physical-layer resources, with the remaining resources allocated to communication tasks.

Building on prior work on localization bounds in wireless networks \cite{shen2010fundamental1,shen2010fundamental,SheDaiWin:J14}, we develop a theoretical framework to derive the \acp{CRLB} of position estimation and the \ac{CRLB} of velocity estimation, leveraging the concept of the \ac{EFIM}. The proposed framework is then employed to quantify the impact of key system parameters, such as network geometry, number of subcarriers, \ac{OFDM} symbols, and receive antennas, on the sensing performance of \ac{ISAC} networks.
The main contributions of this work can be summarized as follows:
\begin{itemize}
\item We study the position and velocity estimation accuracy of \ac{OFDM}-based \ac{ISAC} networks comprising combinations of cooperative monostatic sensors and bistatic \ac{Tx}–\ac{Rx} pairs.
\item We first derive a closed-form expression for the \ac{CRLB} of target position estimation, whose square root defines the \ac{PEB}, for a single monostatic \ac{BS} using the \ac{EFIM}. We then generalize the result to multiple cooperative monostatic \acp{BS} by leveraging the additive property of Fisher information under independent observations.
\item We extend the \ac{PEB} analysis to bistatic configurations by deriving a closed-form expression for a single bistatic \ac{Tx}–\ac{Rx} pair, and generalize the approach to multistatic networks composed of arbitrary sets of bistatic pairs.
\item We derive closed-form expressions for the \ac{VEB}, i.e., the \ac{CRLB} of target velocity estimation, for both cooperative monostatic and multistatic network configurations, explicitly accounting for the influence of position estimation accuracy.
\item We apply the proposed theoretical framework to assess sensing coverage through heatmap visualizations and analyze the impact of key system parameters, including the number of \acp{BS}, available bandwidth, number of antennas, and temporal observations. Furthermore, we show that the framework can be used to identify network configurations that minimize the \ac{PEB} and \ac{VEB} at specific target locations.
\end{itemize}

The remainder of the paper is organized as follows. Section~\ref{sec:intro} introduces the system model, detailing the structure and characteristics of the dual-functional \ac{OFDM} waveform employed for joint sensing and communication. In Section~\ref{sec:pebmonobis}, we derive the \ac{PEB} for both monostatic and bistatic sensing configurations. Section~\ref{sec:netmonoCRLB} presents the derivation of the \ac{PEB} and the \ac{VEB} in a network of cooperative monostatic sensors, while Section~\ref{sec:netmultiCRLB} extends the analysis to multistatic networks. Numerical results validating the theoretical findings are provided in Section~\ref{sec:numres}, while concluding remarks are given in Section~\ref{sec:conclusions}.

Throughout the paper, we use capital boldface letters for matrices and lowercase bold letters for vectors. Additionally, $(\cdot)^\transp$ and $(\cdot)^\herm$ denote transpose and conjugate transpose, respectively; $\bigl\| \cdot \bigr\|$ is the Euclidean norm of vectors; $[\mathbf{X}]_{a:b,c:d}$ denotes a submatrix of $\mathbf{X}$ composed of rows $a$ to $b$ and columns $c$ to $d$; $\mathbf{I}_n$ is the $n \times n$ identity matrix, while $\Tr(\cdot)$ is the trace of a square matrix. $\mathbb{E}\{\cdot\}$ and $\mathbb{V}\{\cdot\}$ represent mean value and variance, respectively. A zero-mean circularly symmetric complex Gaussian random vector with covariance $\boldsymbol{\Sigma}$ is denoted by $\mathbf{x} \thicksim \mathcal{CN}( \mathbf{0},\boldsymbol{\Sigma})$. $|\cdot|$ denotes the absolute value operator, $(\cdot)^*$ denotes complex conjugation, $\Re\{\cdot\}$ and $\Im\{\cdot\}$ the real and imaginary parts, respectively. For a 2D vector $\mathbf{v}$, $|\mathbf{v}|$ denotes its Euclidean norm (i.e., magnitude), and $\angle \mathbf{v}$ denotes its orientation with respect to the reference frame.

\section{System Model for Monostatic and Bistatic ISAC}\label{sec:intro}

\begin{figure*}[t!]
    \centering
    \includegraphics[width=0.43\textwidth]{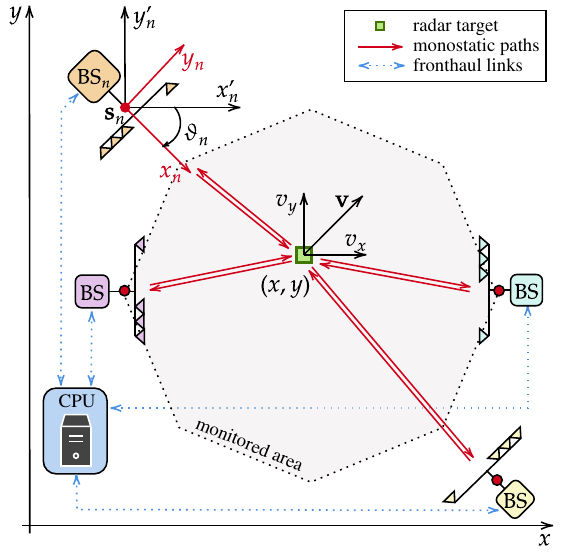} 
    \hspace{7mm}
    \includegraphics[width=0.43\textwidth]{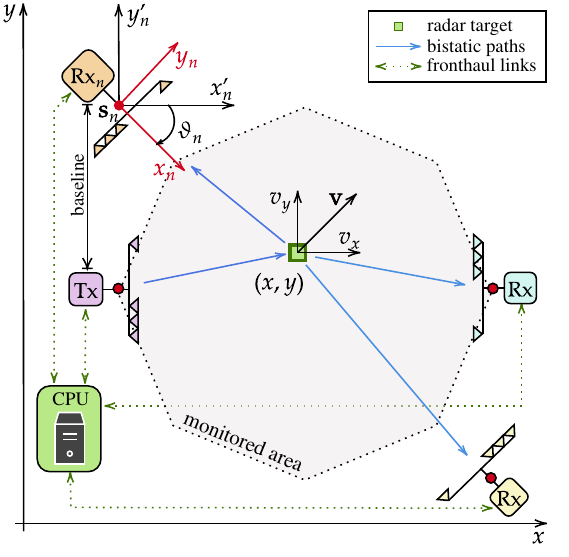}
    \vspace{-0.7em} 
    \begin{minipage}[t]{0.47\textwidth}
        \centering
        \small (a)
    \end{minipage}
    \hfill
    \begin{minipage}[t]{0.47\textwidth}
        \centering
        \small (b)
    \end{minipage}
    \vspace{0.2em} 
    \caption{Target localization scenarios in \ac{ISAC} networks. (a) Network of $\Nbs$ cooperating monostatic \acp{BS}. (b) Multistatic network with one transmitting \ac{BS} and $N_\mathrm{bis}$ receiving \acp{BS}. In both cases, the \acp{BS} are connected to a central \ac{CPU}, which coordinates their operation and performs data fusion.}
    \label{fig:scenarios_combined}
\end{figure*}

The two main types of \ac{ISAC} networks considered in this work are illustrated in Fig.~\ref{fig:scenarios_combined}. These networks are composed of two fundamental building blocks: monostatic sensors and bistatic pairs. For both building blocks, the underlying signal and system models are identical and are described in the following. The manner in which these components are integrated to form cooperative sensing networks is addressed in the subsequent sections.

A multiple-antenna \ac{OFDM} system is considered. The \ac{ISAC} system comprises a transmit antenna array with $N_\mathrm{T}$ elements and a receive antenna array with $N_\mathrm{R}$ elements. The transmit array is employed for both communication and sensing, while the receive array collects echoes from the environment to estimate the target parameters. The \ac{Tx} and \ac{Rx} can either be co-located (i.e., forming a monostatic configuration) or spatially separated (i.e., forming a bistatic configuration).

For both \ac{Tx} and \ac{Rx}, we assume a \ac{ULA} with half-wavelength separation, i.e., $d = \lambda_\mathrm{c}/2$ with $\lambda_\mathrm{c}=c/f_\mathrm{c}$ the wavelength, $c$ the speed of light, and $f_\mathrm{c}$ the carrier frequency.
The system transmits \ac{OFDM} frames with $M_\mathrm{f}$ \ac{OFDM} symbols and $K_\mathrm{a}$ active subcarriers for each frame. The \ac{ELP} representation of the signal transmitted by the $n$th antenna can be written as\footnote{To simplify the notation, the \ac{CP} is omitted in \eqref{eq:base-band}, although it is implicitly included to guarantee robustness against \ac{ISI}. Furthermore, guard subcarriers are included at the spectral edges (outside the $K_\mathrm{a}$ active subcarriers) to limit out-of-band emissions and reduce interference with adjacent channels.}
\begin{equation}
       \label{eq:base-band}
        s_n(t) = \sum_{m=0}^{M_\mathrm{f}-1}\left( \sum_{k=0}^{K_\mathrm{a}-1}x_n[k,m]e^{\imath 2 \pi \frac{k}{T}t}\right)g(t-m\Ts)  
\end{equation}
where $x_n[k,m]$ is the transmitted modulation symbol at the $k\text{th}$ subcarrier in the $m\text{th}$ \ac{OFDM} symbol, taken from a complex modulation alphabet $\mathcal{A}$ and mapped through digital precoding at the $n\text{th}$ transmitting antenna; $g(t)$ is the employed pulse, $\Df = 1/T$ is the subcarrier spacing, and $\Ts=T+T_\mathrm{CP}$ is the \ac{OFDM} symbol duration including the \ac{CP} of duration $T_\mathrm{CP}$. 

\subsection{Transmitted Sensing Signal}
In the \ac{ISAC} system considered in this work, sensing and communication functionalities share the physical layer resources via time and/or frequency division multiplexing over the OFDM time-frequency grid. Specifically, a fraction $\rho_f = K/K_\mathrm{a}$ of contiguous subcarriers (frequency domain) and $\rho_t = M/M_\mathrm{f}$ of \ac{OFDM} contiguous symbols (time domain) within a frame is allocated to sensing operations, where $K$ and $M$ denote the number of subcarriers and \ac{OFDM} symbols assigned to sensing, respectively, as in \cite{FavMatPucPaoXuGio:J25}.

We consider the common case where the same symbol $x[k,m]$ is transmitted to all antennas via beamforming, so the transmit signal vector across antennas, with elements $x_n[k,m]$, is given by $\mathbf{x}[k,m] = \mathbf{w}_\mathrm{T} x[k,m] \in \mathbb{C}^{N_\mathrm{T} \times 1}$, where $\mathbf{w}_\mathrm{T} \in \mathbb{C}^{N_\mathrm{T} \times 1}$ is the beamforming (precoding) vector. Here, $k=0,\dots,K-1$ and $m=0,\dots,M-1$. Without loss of generality, the constellation is normalized so that $\EX{|x[k,m]|^2}=1$.

The vector $\mathbf{w}_\mathrm{T}$ can be optimized according to a specific rule or designed to accommodate multiple beams. For example, it is possible to choose such a vector so that the power of the \ac{OFDM} signal to be transmitted is split between communication and sensing, namely, the total available power is in part exploited to sense the environment and in part directed to the \ac{UE} (see, e.g., \cite{zhang2018multibeam,barneto2020multibeam}).  
In this work, $\mathbf{w}_\mathrm{T}$ is a beamforming vector that points in a specific direction where a radar target is expected. For example, considering a beam steering approach, this can be expressed as
\begin{equation}
    \mathbf{w}_{\text{T}} = \sqrt{\frac{P_\mathrm{avg}}{N_\mathrm{T}}} \mathbf{a}(\theta_\mathrm{T,s})
    \label{eq:BFvector}
\end{equation}
where $P_\mathrm{avg} = P_\mathrm{T}/K_\mathrm{a}$ is the average transmit power per subcarrier, being $P_\mathrm{T}$ the total transmit power of the \ac{OFDM} signal. Moreover, $\mathbf{a}(\theta_\mathrm{T,s}) \in \mathbb{C}^{N_\mathrm{T} \times 1}$ is the steering vector for the sensing direction $\theta_\mathrm{T,s}$. Considering a \ac{ULA} and taking its center as a reference, the spatial steering vector at a given \ac{DoA}/\ac{DoD} $\theta$ is  \cite[Chapter~9]{richards_book}
\begin{equation}
\av(\theta) = \big [e^{-\imath (\Na -1) \pi d \sin \theta / \lambda_\mathrm{c}},\dots, e^{\imath (\Na -1) \pi d \sin \theta / \lambda_\mathrm{c}} \big]^\transp 
\label{eq:steering2}
\end{equation}
where $N_\mathrm{a}$ is the number of array antenna elements. 

\subsection{Received Sensing Signal} \label{sec:rxSignal}
Considering negligible \ac{ICI} and \ac{ISI}, the vector $\rv[k,m] \in \mathbb{C}^{N_\mathrm{R} \times 1}$ of the received symbols after the \ac{FFT} block in the \ac{OFDM} \ac{Rx}, is given by\footnote{This is a shorthand notation for a tensor or 3D array where to keep the notation simple, we spot the vector containing samples along the antennas, so we have one vector for each $(k,m)$ pair. The $(k,m)$ pair represents an element of the OFDM time-frequency grid.}
\begin{equation}
    \mathbf{r}[k,m] = \hchan[k,m]\mathbf{x}[k,m] + \boldsymbol{\nu}[k,m] 
    \label{eq:y}
\end{equation}
where $\hchan[k,m] \in \mathbb{C}^{\Nr \times \Nt}$ is the \ac{MIMO} channel matrix for the $k$th subcarrier at time $m$ and $\boldsymbol{\nu}[k,m]\sim \mathcal{CN}(\mathbf{0},\sigma_\mathrm{N}^2 \mathbf{I}_{\Nr})$ is the \ac{AWGN} at \ac{Rx} antennas with $\sigma_\mathrm{N}^2 = N_0 \Delta f$. Here $N_0 = k_\mathrm{B} T_0 n_\mathrm{F}$ is the one-sided noise \ac{PSD},  being $k_\mathrm{B}$ the Boltzmann constant, $T_0$ the reference temperature, and $n_\mathrm{F}$ the \ac{Rx} noise figure. Note that $\boldsymbol{\nu}[k,m]$ are random vectors \ac{i.i.d.} across the subcarriers and the time.

Considering reflections from $L$ scatterers, the channel matrix can be written as
\begin{equation}\label{eq:channel-matrixFull}
    \hchan[k,m] = \sum_{l = 1}^{L} \underbrace{\bar{\alpha}_l e^{\imath 2\pi m \Ts f_{\mathrm{D},l}}e^{-\imath 2\pi k \Df \tau_l}}_{\triangleq \beta_l} \mathbf{b}(\theta_{\mathrm{R},l})\mathbf{a}^\herm(\theta_{\mathrm{T},l})
\end{equation}
where $\bar{\alpha}_l = \alpha_l e^{j\phi_l}$ is the complex channel gain associated with the $l$th propagation path, incorporating both attenuation and phase shift. According to the radar equation, the power gain $\alpha_l^2$ can be expressed for monostatic and bistatic configurations as \cite{richards_book}
\begin{equation}
\alpha_{l,\mathrm{mono}}^2~= \frac{G_\mathrm{T} G_\mathrm{R} c^2 \sigma_l }{(4 \pi)^3 f^2_\mathrm{c}\, r_l^4},
 \qquad
\alpha_{l,\mathrm{bis}}^2~= \frac{G_\mathrm{T} G_\mathrm{R} c^2 \sigma_l }{(4 \pi)^3 f^2_\mathrm{c}\, r_{\mathrm{T},l}^2 r_{l,\mathrm{R}}^2}.
\label{eq:alpha} 
\end{equation}
Here, $G_\mathrm{T}$ and $G_\mathrm{R}$ denote the gains of the \ac{Tx} and \ac{Rx} antenna elements, respectively. The distance between the monostatic system and the $l$th scatterer is $r_l$, while $r_{\mathrm{T},l}$ and $r_{l,\mathrm{R}}$ represent the distances from the \ac{Tx} to the scatterer $l$  and from the scatterer to the \ac{Rx}, respectively, in a bistatic configuration. The parameter $\sigma_l$ denotes the \ac{RCS} of the $l$th scatterer.

The terms $\tau_l$, $f_{\mathrm{D},l}$, $\theta_{\mathrm{R},l}$ and $\theta_{\mathrm{T},l}$ in \eqref{eq:channel-matrixFull} are the propagation delay, the Doppler shift, the \ac{DoA} and \ac{DoD} associated with the $l$th path. The vectors $\mathbf{b}(\theta_{\mathrm{R},l})$ and $\mathbf{a}(\theta_{\mathrm{T},l})$ are the array response vectors at the \ac{Rx} and \ac{Tx}, respectively. According to \eqref{eq:steering2}, and considering typical half-wavelength inter-element spacing, these are given, respectively, by%
\begin{align}
\av(\thetat)&= \left[e^{-\imath \frac{\Nt-1}{2}\pi \sin \thetat},
\dots,e^{\imath \frac{\Nt-1}{2}\pi \sin \thetat}\right]^\transp \nonumber \\
\bv(\thetar)&= \left[e^{-\imath \frac{\Nr-1}{2}\pi \sin \thetar},
\dots,e^{\imath \frac{\Nr-1}{2}\pi \sin \thetar}\right]^\transp 
\end{align}
%
from which it can be readily noticed that
\begin{equation}\label{eq:steerprop}
\av^\herm(\thetat)\av(\thetat)=\Nt, \quad
\bv^\herm(\thetar)\bv(\thetar)=\Nr.
\end{equation}
Moreover, given $a_i$ and $b_i$ the $i$-th elements of $\av(\thetat)$ and $\bv(\thetar)$, respectively, the derivatives with respect to $\thetat$ and $\thetar$ are
\begin{align}\label{eq:deriv}
\ap(\thetat) & \triangleq  \frac{d \av(\thetat)}{d\thetat} = \left[-\imath a_1\frac{\Nt-1}{2}\pi \cos \thetat, \right.\\
&\left. -\imath a_2\frac{\Nt-3}{2}\pi \cos \thetat,\dots, \imath a_{\Nt}\frac{\Nt-1}{2}\pi \cos \thetat\right]^\transp \nonumber\\
\bp(\thetar) & \triangleq  \frac{d \bv(\thetar)}{d\thetar} = \left[-\imath b_1\frac{\Nr-1}{2}\pi \cos \thetar, \right.\nonumber\\
& \left.-\imath b_2\frac{\Nr-3}{2}\pi \cos \thetar,\dots, \imath b_{\Nr}\frac{\Nr-1}{2}\pi \cos \thetar\right]^\transp \nonumber
\end{align}
which leads to the orthogonality property between steering vectors and their derivatives \cite{BekTab:J06}
\begin{equation}\label{eq:orthogprop}
\ap^\herm(\thetat) \av(\thetat)=\bp^\herm(\thetar) \bv(\thetar)=0.
\end{equation}
According to \eqref{eq:deriv}, we define the second derivative of the vector $\bv(\thetar)$ with $\bpp(\thetar)\triangleq d^2 \bv(\thetar)/d \thetar^2$. A useful property of $\bv(\thetar)$ is the following
\begin{align}\label{eq:bppprop}
&\bh(\thetar)\bpp(\thetar)=\sum_{l=-(\Nr-1)/2}^{(\Nr-1)/2} b_l^*(\thetar)\Ddot{b}_l(\thetar)\\
&=-\pi^2\cos^2(\thetar)\!\!\!\sum_{l=-(\Nr-1)/2}^{(\Nr-1)/2} \!\!\! l^2=-\frac{\pi^2}{12}(\Nr^2-1)\Nr \cos^2(\thetar).\nonumber
\end{align}

Without loss of generality, we focus on a single-scatterer scenario, i.e., $L=1$, so that the index $l$ can be dropped and the channel matrix in \eqref{eq:channel-matrixFull} simplifies to
\begin{equation}\label{eq:channel-matrix}
    \hchan[k,m] = \alpha e^{\imath \phi} e^{\imath 2\pi m \Ts f_{\mathrm{D}}}e^{-\imath 2\pi k \Df \tau} \mathbf{b}(\theta_{\mathrm{R}})\mathbf{a}^\herm(\theta_{\mathrm{T}})
\end{equation}
and the corresponding received signal reduces to
\begin{align}
    \mathbf{r}[k,m] = &\alpha  e^{\imath \phi} e^{\imath 2\pi m \Ts f_{\mathrm{D}}}e^{-\imath 2\pi k \Df \tau} \mathbf{b}(\theta_{\mathrm{R}})\mathbf{a}^\herm(\theta_{\mathrm{T}})\mathbf{w}_\mathrm{T} x[k,m]\nonumber\\ &+ \boldsymbol{\nu}[k,m]. 
    \label{eq:y2}
\end{align}

After element-wise division, the transmitted symbols $x[k,m]$ (assumed known) are removed as nuisance terms, yielding the following processed received samples \cite{Braun,Col:J22,PucPaoGio:J22}\footnote{This approach is often referred to as reciprocal filtering.}
\begin{align}\label{eq:recsigdiv}
    \y[k,m]=&\alpha  e^{\imath \phi}e^{\imath 2\pi m \Ts f_{\mathrm{D}}}e^{-\imath 2\pi k \Df \tau}  \mathbf{b}(\theta_{\mathrm{R}})\mathbf{a}^\herm(\theta_{\mathrm{T}})\mathbf{w}_\mathrm{T}  \nonumber\\ 
    &+ \mathbf{n}[k,m]
\end{align}
%
where $\mathbf{n}[k,m] = \boldsymbol{\nu}[k,m]/x[k,m]\sim \mathcal{CN}(\mathbf{0},\widetilde{\sigma}_\mathrm{N}^2  \mathbf{I}_{\Nr})$ with 
\begin{equation}
\widetilde{\sigma}_\mathrm{N}^2= \mathbb{E}\left\{\frac{|\nu[k,m]|^2}{|x[k,m]|^2}\right\} = \sigma_\mathrm{N}^2\cdot \underbrace{\mathbb{E}\left\{\frac{1}{|x[k,m]|^2}\right\}}_{\eta}. 
\label{eq:noise_variance}
\end{equation}

\begin{figure*}[ht]
\setcounter{MYtempeqncnt}{\value{equation}}
\begin{equation}\label{eq:likelihood}
f(\boldsymbol{\mathcal Y})=\prod_{k=0}^{K-1} \prod_{m=0}^{M-1} \frac{1}{(\pi \widetilde{\sigma}_\mathrm{N}^2)^{\Nr}}\exp\biggl(-\frac{1}{\widetilde{\sigma}_\mathrm{N}^2}\Bigl\|\y[k,m]-\alpha  e^{\imath \phi}e^{\imath 2\pi m \Ts f_{\mathrm{D}}}e^{-\imath 2\pi k \Df \tau}  \mathbf{b}(\theta_{\mathrm{R}})\mathbf{a}^\herm(\theta_{\mathrm{T}})\mathbf{w}_\mathrm{T}\Bigr\|^2\biggr)
\end{equation}
\hrulefill
\end{figure*}

Recalling that the noise after \ac{FFT} blocks is white across the time, frequency, and spatial (antenna) domain, the likelihood function of the received ensemble  
\begin{equation}
\boldsymbol{\mathcal Y}=\{\y[k,m]\}_{k=0,\dots,K-1,m=0,\dots,M-1}
\label{eq:rx_ens}
\end{equation}
is expressed as in \eqref{eq:likelihood} at the top of the next page.
%
%
The corresponding log-likelihood is then given by
\begin{align}\label{eq:LLF}
\ln f(\boldsymbol{\mathcal Y})=-\sum_{k=0}^{K-1} \sum_{m=0}^{M-1}& \Nr \ln \widetilde{\sigma}_\mathrm{N}^2 \\
+\frac{1}{\widetilde{\sigma}_\mathrm{N}^2}\Bigl\|\y[k,m]&-\alpha  e^{\imath \phi} e^{\imath 2\pi m \Ts f_{\mathrm{D}}}e^{-\imath 2\pi k \Df \tau}  \h\Bigr\|^2\nonumber
\end{align}
where terms not relevant for the derivation of the \ac{FIM} have been omitted. The vector
\begin{equation}
\h= 
\bv(\thetar) \gamma \quad \textrm{with} \quad\gamma=\mathbf{a}^\herm(\theta_{\mathrm{T}})
\mathbf{w}_\mathrm{T}
\end{equation}
captures the spatial dimension of the sensing channel.

\subsubsection{Remark on the SNR}
From \eqref{eq:y2}, we define the \ac{SNR}, for the given target, per receiving antenna as
\begin{equation}
    \SNR=\frac{\alpha^2 |\mathbf{w}_\mathrm{T}^\herm\mathbf{a}(\theta_\mathrm{T})|^2}{\sigma^2_\mathrm{N}}\mathbb{E}\Bigl\{\bigl|x[k,m]\bigr|^2\Bigr\}=\frac{\alpha^2 |\gamma|^2}{\sigma^2_\mathrm{N}}.
    \label{eq:SNR}
\end{equation}
If $\mathbf{w}_\mathrm{T}$ is chosen according to \eqref{eq:BFvector}, then
\begin{equation}
|\gamma|^2 = \frac{P_\mathrm{avg}\left|\Upsilon(\theta_\mathrm{T},\theta_\mathrm{T,s})\right|^2}{\Nt}
\label{eq:gamma2}
\end{equation}
where $\Upsilon(\theta_\mathrm{T},\theta_\mathrm{T,s})=\mathbf{a}^\herm(\theta_\mathrm{T})\mathbf{a}(\theta_\mathrm{T,s})$ represents the beamforming gain. When $\theta_\mathrm{T,s} = \theta_\mathrm{T}$, i.e., when pointing towards the target, using \eqref{eq:steerprop}, we have $\left|\Upsilon(\theta_\mathrm{T},\theta_\mathrm{T,s})\right|^2 = N^2_\mathrm{T}$. Thus, \eqref{eq:gamma2} becomes, $|\gamma|^2=P_\mathrm{avg} \Nt$ and, considering also \eqref{eq:alpha}, the \ac{SNR} in \eqref{eq:SNR} can be rewritten as
%
\begin{equation}
\label{eq:SNR3}  
\SNR=\frac{P_\mathrm{avg}G_\mathrm{T} G_\mathrm{R} N_\mathrm{T} c^2 \sigma}{(4 \pi)^3 f^2_\mathrm{c}\, r^4\sigma^2_\mathrm{N}}.
\end{equation}
Moreover, considering \eqref{eq:noise_variance}, we define the \ac{SNR} related to \eqref{eq:recsigdiv}, i.e., after performing element-wise division with the transmitted symbols $x[k,m]$, as follows
\begin{equation}
    \SNR_\mathrm{post-div}=\frac{\alpha^2 |\mathbf{w}_\mathrm{T}^\herm\mathbf{a}(\theta_\mathrm{T})|^2}{\widetilde{\sigma}^2_\mathrm{N}}=\frac{\alpha^2 |\gamma|^2}{\eta\,\sigma^2_\mathrm{N}}=\frac{\mathrm{SNR}}{\eta}.
    \label{eq:SNR_postdiv}
\end{equation}
Hence, the term $\alpha^2 \|\h\|^2/\widetilde{\sigma}_\mathrm{N}^2$ in the \ac{FIM}
can be simplified using \eqref{eq:steerprop} and  \eqref{eq:SNR_postdiv} as
\begin{align}
    \frac{\alpha^2\|\h\|^2}{\widetilde{\sigma}_\mathrm{N}^2}&=\frac{\alpha^2(\mathbf{b}(\theta_\mathrm{R})\mathbf{a}^\herm(\theta_\mathrm{T})\mathbf{w}_\mathrm{T})^\herm \mathbf{b}(\theta_\mathrm{R})\mathbf{a}^\herm(\theta_\mathrm{T})\mathbf{w}_\mathrm{T}}{\widetilde{\sigma}^2_\mathrm{N}}\nonumber\\
    &=\frac{\alpha^2N_\mathrm{R}\mathbf{w}_\mathrm{T}^\herm\mathbf{a}(\theta_\mathrm{T})\mathbf{a}^\herm(\theta_\mathrm{T})\mathbf{w}_\mathrm{T}}{\widetilde{\sigma}^2_\mathrm{N}}\nonumber\\
    &=\frac{\alpha^2 N_\mathrm{R}|\gamma|^2}{\widetilde{\sigma}^2_\mathrm{N}}=\frac{\Nr \SNR}{\eta}.
\end{align}
%
%

\subsubsection{SNR penalty due to the constellation shape}
As evident from \eqref{eq:SNR_postdiv}, the sensing \ac{SNR} is reduced by a factor $\eta$, defined in \eqref{eq:noise_variance} and given by
\begin{equation}
    \eta = \frac{1}{P} \sum_{i=1}^{P} \frac{1}{|x_i|^2}
    \label{eq:SNR_factor}
\end{equation}
where $x_i$ denotes the $i$th constellation point in a modulation scheme with $P$ equiprobable symbols.

When a constant-envelope modulation is employed (e.g., QPSK), all constellation points satisfy $|x_i|^2 = 1$, and thus $\eta = 1$, implying no penalty. In contrast, for modulations with non-constant envelope (e.g., M-QAM), we have $\eta > 1$, which leads to an increase in the effective noise variance, i.e., $\widetilde{\sigma}_\mathrm{N}^2 > \sigma_\mathrm{N}^2$, after element-wise symbol division. This results in a degradation of the effective sensing \ac{SNR}.

For example, $\eta \approx 1.89$ for 16-QAM, $\eta \approx 2.69$ for 64-QAM, and $\eta \approx 3.44$ for 256-QAM, corresponding to an \ac{SNR} loss of approximately $2.76\,\mathrm{dB}$, $4.29\,\mathrm{dB}$, and $5.36\,\mathrm{dB}$, respectively, when compared to a constant-envelope signal. These findings are consistent with recent results indicating that constant-envelope waveforms are more favorable for sensing applications~\cite{XioLiu:J23,KesMojLacetal:J25}.

\section{Position Error Bound for Monostatic and Bistatic Sensing} \label{sec:pebmonobis}
\subsection{Fisher Information Matrix and CRLB}\label{sec:CRLB}
The \ac{FIM} of the vector of parameters ${\boldsymbol{\Theta}}=
[\alpha,\phi,\fd,\tau,\theta_\mathrm{R}]^\transp$ is calculated from \eqref{eq:LLF} as
\begin{equation}\label{eq:FIMdef}
    [\I(\boldsymbol{\Theta})]_{i,j}=-\EX{\frac{\partial^2 \ln f(\boldsymbol{\mathcal Y})}{\partial \theta_i \partial\theta_j}} \quad i,j\in \{1,\dots, 5\}.
\end{equation}

Such a matrix is derived in Appendix \ref{sec:appA} and reported in \eqref{eq:FIMofdm} on the top of the next page.

By inverting \eqref{eq:FIMofdm} at the top of the next page, we obtain the expression of the \ac{CRLB} for each parameter as 
\begin{figure*}
\setlength{\arraycolsep}{2pt}
\begin{align}\label{eq:FIMofdm}
\I(\boldsymbol{\Theta})= K M \alpha^2 \|\h\|^2 & /\widetilde{\sigma}_\mathrm{N}^2\\
&
\times\begin{bmatrix}
\frac{2}{\alpha^2} & 0 & 0 & 0 & 0\\
0 & 2 & 2\pi  \Ts (M-1)
 & -2\pi \Df (K-1)
& 0\\
0 & 2\pi \Ts (M-1)
 & \frac{4 \pi^2 \Ts^2 (2M-1)(M-1)}{3} & -2\pi^2 \Ts \Df (M-1)(K-1) & 0\\
0 & -2\pi \Df (K-1) & -2\pi^2 \Ts \Df (M-1)(K-1) & \frac{4\pi^2 \Df^2 (2K-1)(K-1)}{3} & 0\\
0 & 0 & 0 & 0 & \frac{\pi^2 (\Nr^2-1)\cos^2(\thetar)}{6}
\end{bmatrix}\nonumber
\end{align}
\hrulefill
\vspace*{4pt}
\end{figure*}
\begin{align}\label{eq:IFIM}
  \CRB(\alpha)&=[\I^{-1}]_{1,1}=
  \frac{\alpha ^2 \, \eta}{2 K M \Nr \SNR}\\
  \CRB(\phi)&=[\I^{-1}]_{2,2}=
  \frac{(7 K M+K+M-5)\, \eta}{2 (K^2+K) \left(M^2+M\right)
   \Nr \SNR}
\nonumber\\
\CRB(\fd)&=[\I^{-1}]_{3,3}=\frac{3 \, \eta}{2 \pi^2 \Ts^2 K M (M^2-1) \Nr
   \SNR}
\nonumber\\
  \CRB(\tau)&=[\I^{-1}]_{4,4}=
  \frac{3 \, \eta}{2 \pi ^2 \Delta f^2 M K (K^2-1)\Nr \SNR} 
\nonumber\\
  \CRB(\theta_\mathrm{R})&=[\I^{-1}]_{5,5}=
  \frac{6 \, \eta}{\pi^2 K M (\Nr^2-1) \Nr \SNR\cos^2(\thetar)}. \nonumber
\end{align}

\subsubsection{Range estimation}
Recalling that the range is $r=c\tau /2$ and the bistatic range is $\bar{r}= r_\mathrm{T} + r_\mathrm{R} = c \tau$, with $r_\mathrm{T}$ the \ac{Tx}-target distance and $r_\mathrm{R}$ the target-\ac{Rx} distance, by the scaling law property of variance we obtain
\begin{equation}\label{eq:crlbrange}
\CRB(r)=\left(\frac{c}{2}\right)^2\frac{3\, \eta}{2 \pi ^2 \Delta f^2 M K (K^2-1)\Nr \SNR} 
\end{equation}
and $\CRB(\bar{r})=4\,\CRB(r)$. 
%


\subsection{Position Error Bound for a Monostatic Sensor} \label{sec:monoCRLB}
Consider a given monostatic \ac{ISAC} \ac{BS}, indicated by the index $n$. Hereinafter, we indicate with subscript $n$ the target parameters related to the \ac{BS} $n$ and the received ensemble in \eqref{eq:rx_ens} as $\boldsymbol{\mathcal{Y}}_n$. Let $\mathbf{s}_n=[\xbs,\ybs]^\transp$ and $\p = [x,y]^\transp$ denote the positions of the monostatic \ac{BS} and the point-like target, respectively, in the global reference frame. Furthermore, let $\mathbf{p}_n = [x_n, y_n]^\transp$ represent the position of the target in the local reference frame of \ac{BS} $n$, which is centered at $\mathbf{s}_n$ with the local x-axis rotated by angle $\vartheta_n$ (measured from the global x-axis to the local x-axis, in the interval $(-\pi, \pi]$). The relationship between $\p$ and $\p_n$ is given by
\begin{equation}
\begin{cases}
x_n = x_n'\cos(\thetan) + y_n'\sin(\thetan) \\
y_n = -x_n'\sin(\thetan) + y_n'\cos(\thetan) 
\end{cases}
\label{eq:global_to_local}
\end{equation}
where $x_n'=x-\xbs$ and $y_n'=y-\ybs$.
The true target position in the local reference system can be expressed as $\p_n = [x_n, y_n]^\transp = [r_n \cos(\thetarn), r_n \sin(\thetarn)]^\transp$, and the estimated position is denoted by $\ph_n = [\widehat{x}, \widehat{y}]^\transp = [\widehat{r}_n \cos(\widehat{\theta}_{\mathrm{R},n}), \widehat{r}_n \sin(\widehat{\theta}_{\mathrm{R},n})]^\transp$. Here, $r_n = \tfrac{c\tau_n}{2} = \sqrt{x_n^2 + y_n^2}$ is the distance between the target and the \ac{BS} $n$, $\thetarn = \arctan{(y_n/x_n)} \in (-\pi/2,\pi/2]$ is the target \ac{DoA}, while their estimates are indicated by $\widehat{r}_n$ and $\widehat{\theta}_{\mathrm{R},n}$, respectively.

\begin{figure*}[!t]
\setcounter{MYtempeqncnt}{\value{equation}}
\setlength{\arraycolsep}{2pt}
\begin{equation}
\CRB(\p_n)= \frac{6 \, \eta}{\pi^2 K M \Nr \SNR^{(n)}}\left[\frac{c^2/16}{\Df^2 (K^2-1)}+\frac{r_n^2}{(\Nr^2-1) \cos^2(\thetarn)}\right].
\label{eq:CRLBp_mono_exp}
\end{equation}
\hrulefill
\vspace*{4pt}
\end{figure*}

The position estimation error is defined as the Euclidean distance between the true and estimated target position, i.e., $e_{\mathbf{p}_n} = \|\ph_n - \p_n\| = \sqrt{(x_n - \widehat{x}_n)^2 + (y_n - \widehat{y}_n)^2}$. The \ac{CRLB} for target position estimation in a monostatic sensing scenario can be computed by retaining only the components of the \ac{FIM} \eqref{eq:FIMofdm} associated with the target parameters $\tau_n$ and $\thetarn$ at \ac{BS} $n$, through the use of the \ac{EFIM} \cite{shen2010fundamental}. Given the parameter vector ${\boldsymbol{\Theta}_n}=[\alpha_n,\phi_n,\fdn,\tau_n,\theta_{\mathrm{R},n}]^\transp$ we partition it into ${\boldsymbol{\Theta}_n}=[\Thetanone^\transp,\Thetantwo^\transp]^\transp$, with $\Thetanone=[\alpha_n,\phi_n,\fdn]^\transp$ and $\Thetantwo=[\tau_n,\theta_{\mathrm{R},n}]^\transp$,
which induces the partition of the \ac{FIM} in \eqref{eq:FIMofdm} as
\begin{equation}\label{eq:FIMpart}
\I(\boldsymbol{\Theta}_n)=
\begin{bmatrix}
\mathbf{A} & \mathbf{B} \\
\mathbf{B}^\transp & \mathbf{C}
\end{bmatrix}
\end{equation}
where $\mathbf{A}\in \mathbb{R}^{3\times 3}$, $\mathbf{B}\in \mathbb{R}^{3\times 2}$, and $\mathbf{C}\in \mathbb{R}^{2\times 2}$.
Then, the \ac{EFIM} is given by
\begin{align}\label{eq:EFIMdef}
\Ie(\Thetantwo) & = \mathbf{C}-\mathbf{B}^\transp\mathbf{A}^{-1}\mathbf{B}  \\
& = \SNR^{(n)}\,\Nr K M/\eta \nonumber \\
& \qquad \times
\begin{bmatrix}
\frac{2\pi^2\Df^2 (K^2-1)}{3} & 0 \\
0 & \frac{\pi^2(\Nr^2-1)\cos^2(\thetarn)}{6}\nonumber
\end{bmatrix}
\end{align}
which corresponds to the Schur complement of $\mathbf{A}$.\footnote{A key property of the \ac{EFIM} is that $$[\I^{-1}(\boldsymbol{\Theta}_n)]_{4:5,4:5}=\Ie^{-1}(\Thetantwo)$$ which implies that the \ac{EFIM} retains all the necessary information to derive the information inequality for $\Thetantwo$.} In \eqref{eq:EFIMdef}, $\SNR^{(n)}$ denotes the \ac{SNR} of the target echo received at \ac{BS} $n$. According to \eqref{eq:SNR3}, this quantity depends on several parameters, including the target \ac{RCS} (as seen by \ac{BS} $n$) and the range $r_n$.
Since we are interested in the \ac{CRLB} of position estimation, a reparameterization is required to compute the \ac{EFIM} \eqref{eq:EFIMdef} for $\p_n$, i.e.,
\begin{equation}
\label{eq:EFIMpdef}
[\Ie(\p_n)]_{i,j}=-\EX{\frac{\partial^2 \ln f(\boldsymbol{\mathcal Y}_n)}{\partial p_{n,i} \,\partial p_{n,j}}} \quad i,j\in \{1,2\}. 
\end{equation}
This can be obtained by applying the transformation 
\begin{equation}\label{eq:EFIM_p_mono}
\Ie(\p_n)=\Jm^\transp \Ie(\Thetantwo)\Jm
\end{equation}
where $\Jm$ is the Jacobian of the transformation $\{\thetarn=\arctan(y_n/x_n),\tau_n=\frac{2}{c}\sqrt{x_n^2+y_n^2}\}$ which corresponds to 
\begin{equation}
\Jm=
\begin{bmatrix}
\frac{\partial \tau_n}{\partial x_n} & \frac{\partial \tau_n}{\partial y_n} \\
\frac{\partial \thetarn}{\partial x_n} & \frac{\partial \thetarn}{\partial y_n}
\end{bmatrix}
=
\begin{bmatrix}
\frac{2}{c}\frac{x_n}{\sqrt{x_n^2+y_n^2}} & \frac{2}{c}\frac{y_n}{\sqrt{x_n^2+y_n^2}} \\
-\frac{y_n}{x_n^2+y_n^2} & \frac{x_n}{x_n^2+y_n^2}
\end{bmatrix}.
\end{equation}

Hence, the \ac{CRLB} is 
\begin{equation}\label{eq:CRLB}
\mathbb{V}\{\widehat{\mathbf{p}}_n\}\geq\CRB(\p_n)=\Tr(\Ie^{-1}(\p_n))
\end{equation}
and, using the cyclic property of the trace, it can be further elaborated as
\begin{align}
\CRB(\p_n)&=\Tr(\Ie^{-1}(\p_n))=\Tr\Bigl(\Jm^{-1}\Ie^{-1}(\Thetatwo)\bigl(\Jm^{-1}\bigr)^\transp\Bigr)\nonumber\\
&=\Tr\Bigl(\Ie^{-1}(\Thetantwo)\bigl(\Jm^{-1}\bigr)^\transp\Jm^{-1}\Bigr)\nonumber \\ 
&=\Tr\Bigl(\Ie^{-1}(\Thetantwo)\mathbf{M}_\Xi\Bigr)
\end{align}
where $\mathbf{M}_\Xi=\diag(c^2/4,x_n^2+y_n^2)$. Therefore,
\begin{equation}
\CRB(\p_n)=\frac{c^2}{4}[\Ie^{-1}(\boldsymbol{\Theta}_{n,2})]_{1,1}+r_n^2[\Ie^{-1}(\boldsymbol{\Theta}_{n,2})]_{2,2}.
\label{eq:CRLBp_mono}
\end{equation}
By computing the inverse of $\Ie(\Thetantwo)$ in \eqref{eq:EFIMdef} and replacing in \eqref{eq:CRLBp_mono} the terms on its main diagonal, we obtain the expression in \eqref{eq:CRLBp_mono_exp} at the top of this page for the position estimation \ac{CRLB}.
Then, the \ac{PEB} is defined as
\begin{equation}
    \mathrm{PEB} = \sqrt{\CRB(\p_n)}.
\label{eq:PEB}
\end{equation}

\subsection{Position Error Bound for a Bistatic Pair} \label{sec:bis CRLB}
Consider an \ac{ISAC} system where the \ac{Tx} and \ac{Rx} are not co-located, i.e., a bistatic configuration. Following the approach in Section~\ref{sec:monoCRLB}, we introduce an index $n$ to denote the target parameters associated with \ac{Rx} $n$.

In a global reference frame, we indicate with $\mathbf{s}_\mathrm{T} = [x_\mathrm{T}, y_\mathrm{T}]^\transp$, $\mathbf{s}_n = [x^{(n)}_\mathrm{R}, y^{(n)}_\mathrm{R}]^\transp$ and $\p = [x, y]^\transp$, the positions of the \ac{Tx}, the \ac{Rx} $n$, and the target, respectively.
Moreover, $\p_n = [x_n, y_n]^\transp = [r_n \cos(\thetarn), r_n\sin(\thetarn)]^\transp$ is the target position in the local reference frame centered at $\mathbf{s}_n$ and oriented by an angle $\vartheta_n \in (-\pi,\pi]$, with $r_n$ and $\thetarn$ defined as in Section~\ref{sec:monoCRLB}. 
The relationship between $\p$ and $\p_n$ is given according to \eqref{eq:global_to_local}, with $x'_n = x - x^{(n)}_\mathrm{R}$ and $y'_n = y - y^{(n)}_\mathrm{R}$.

As previously mentioned, in a bistatic radar configuration, the propagation time $\tau_n$ of the signal scattered by the target is related to the sum of the distance between the \ac{Tx} and the target, $r_\mathrm{T}$, and the distance between the target and the \ac{Rx}, $r_n$, via the bistatic range $\bar{r}_n = r_\mathrm{T}+ r_n = \tau_n \cdot c$ \cite{willis05}.
After estimating $\bar{r}_n$ via $\tau_n$, the target can be located on an ellipse with a major axis equal to $\bar{r}_n$ and foci at \ac{Tx} and \ac{Rx} $n$ positions.
The \ac{Tx}, \ac{Rx} $n$, and target form a triangle with base $l_n$  (the distance between \ac{Tx} and \ac{Rx}) called the baseline.

Estimating the \ac{DoA} $\thetarn$ of the reflected echo at the \ac{Rx} $n$, it is possible to determine the distance $r_n$ as \cite{willis05}
\begin{align} \label{eq:Rbis}
    r_n & = \frac{\bar{r}_n^2-l_n^2}{2\bigl(\bar{r}_n-l_n\cos{(\thetarn + \theta_{\mathrm{shift},n})}\bigr)} \nonumber\\
    & = \frac{c^2\tau_n^2-l_n^2}{2\bigl(c \tau_n-l_n\cos{(\thetarn+\theta_{\mathrm{shift},n})}\bigr)}
\end{align}
where $\theta_{\mathrm{shift},n} = \vartheta_n - \beta_n$, with $\beta_n = \arctan\Big(\tfrac{y_\mathrm{T}-y^{(n)}_\mathrm{R}}{x_\mathrm{T}-x^{(n)}_\mathrm{R}}\Big) \in (-\pi,\pi]$.
Then, the components of target position $\p_n$ can be computed as follows
\begin{equation}\label{eq:xybis}
\begin{cases}
x_n = r_n  \cos(\thetarn)\\
y_n = r_n \sin(\thetarn).
\end{cases}
\end{equation}
From \eqref{eq:xybis} the inverse transformation $\Xi^{-1}$ can be defined as $\Xi^{-1}: (\tau_n, \thetarn)\rightarrow (x_n,y_n)$ and its Jacobian $\Jinvm$ is given by 
\begin{equation}\label{eq:Jinvm}
\Jinvm=
\begin{bmatrix}
\frac{\partial x_n}{\partial \tau_n} & \frac{\partial x_n}{\partial \thetarn} \\
\frac{\partial y_n}{\partial \tau_n} & \frac{\partial y_n}{\partial \thetarn}
\end{bmatrix}.
\end{equation}

By replacing \eqref{eq:Rbis} in \eqref{eq:xybis} and introducing the angle $\thetaln = \thetarn + \theta_{\mathrm{shift},n}$, the partial derivatives in \eqref{eq:Jinvm} are given by
\begin{align}
 \frac{\partial x_n}{\partial \tau_n} & = \frac{c \cos(\thetarn)\bigl(l_n^2 + \bar{r}_n^2 -2\, l_n\, \bar{r}_n\cos(\thetaln)\bigr)}{2\bigl(\bar{r}_n -l_n \cos(\thetaln)\bigr)^2} 
\nonumber\\
 \frac{\partial x_n}{\partial \thetarn} & = \frac{\bigl(l_n^2 - \bar{r}_n^2\bigr)\bigl(\bar{r}_n \sin(\thetarn) + l_n \sin(\theta_{\mathrm{shift},n})\bigr)}{2\bigl(\bar{r}_n -l_n \cos(\thetaln)\bigr)^2} 
   \nonumber\\
\frac{\partial y_n}{\partial \tau_n} & = \frac{c \sin(\thetarn)\bigl(l_n^2 +\bar{r}_n^2 -2\, l_n\, \bar{r}_n\cos(\thetaln)\bigr)}{2\bigl(\bar{r}_n -l_n \cos(\thetaln)\bigr)^2} 
\nonumber\\
\frac{\partial y_n}{\partial \thetarn} & = \frac{\bigl(l_n^2 - \bar{r}_n^2\bigr)\bigl(l_n\cos(\theta_{\mathrm{shift},n})-\bar{r}_n\cos(\thetarn)\bigr)}{2\bigl(\bar{r}_n -l_n \cos(\thetaln)\bigr)^2}.
\end{align}

\subsubsection*{Property 1}
\label{prop:inverse_jacobian}
Let $\Xi: \mathbb{R}^2 \rightarrow \mathbb{R}^2$ be a differentiable and locally invertible mapping, such that $\mathrm{\Xi}(x_n, y_n) = (\tau_n, \thetarn)$. Let $\Jm$ denote the Jacobian of $\Xi$, and $\Jinvm$ the Jacobian of its inverse. Then, the following identity holds
\[
\Jinvm = \Jm^{-1}.
\]

\begin{figure*}[!t]
\setcounter{MYtempeqncnt}{\value{equation}}
\setlength{\arraycolsep}{2pt}
\begin{equation}
\CRB^{\mathrm{bis}}(\p_n)= \frac{3\,\eta\bigl(l_n^2 + \bar{r}_n^2 - 2\,l_n\,\bar{r}_n \cos(\thetaln)\bigr)}{8 \pi^2\, \SNR^{(n)}\, \Nr K M \bigl(\bar{r}_n - l_n \cos(\thetaln)\bigr)^4}\left[\frac{c^2\bigl(l^2 + \bar{r}_n^2 - 2\,l_n\, \bar{r}_n \cos(\thetaln)\bigr)}{\Df^2 (K^2-1)}+\frac{4(l_n^2 - \bar{r}_n^2)^2}{(\Nr^2-1)\cos^2(\thetarn)}\right].
\label{eq:CRLBp_bis_exp}
\end{equation}
\hrulefill
\vspace*{4pt}
\end{figure*}

Let $\Ie(\p_n)$ be the \ac{EFIM} for $\p_n$. Using \textit{Property 1} inside \eqref{eq:EFIM_p_mono}, this is given by
\begin{equation}
    \label{eq:EFIM_p_bis}
    \Ie(\p_n)=(\Jinvm^{-1})^\transp \Ie(\Thetantwo)\Jinvm^{-1}
\end{equation}
The \ac{CRLB} of position estimation in a bistatic system is then given by
\begin{align}\label{eq:CRLB_bis}
\CRB_{\mathrm{bis}}(\p_n)&=\Tr(\Ie^{-1}(\p_n))\nonumber\\
&=\Tr(\Jm^{-1}\Ie^{-1}(\Thetantwo)[\Jm^{-1}]^\transp)\nonumber\\
& =\Tr(\Ie^{-1}(\Thetantwo)[\Jinvm]^\transp\Jinvm)\nonumber\\
& =\Tr\left(\Ie^{-1}(\Thetantwo)\mathbf{M}_{\Xi^{-1}}\right)
\end{align}
where $\Ie(\Thetantwo)$ is given in \eqref{eq:EFIMdef}, $\mathbf{M}_{\mathrm{\Xi^{-1}}}=[a_{1,1}, a_{1,2};\, a_{2,1}, a_{2,2}]$, with
\begin{align}\label{eq:invM}
a_{1,1} & = \frac{c^2 \bigl(l_n^2 + \bar{r}_n^2 - 2\,l_n\,\bar{r}_n \cos(\thetaln)\bigr)^2}{4\bigl(\bar{r}_n - l_n \cos(\thetaln)\bigr)^4} \nonumber \\
a_{2,2} & = \frac{(l_n^2 - \bar{r}_n^2)\bigl(l_n^2 + \bar{r}_n^2 -2\,l_n\,\bar{r}_n \cos(\thetaln)\bigr)}{4\bigl(\bar{r}_n - l_n \cos(\thetaln)\bigr)^4}.
\end{align}
Therefore, because of the diagonal nature of $\Ie^{-1}(\Thetantwo)$ (see \eqref{eq:EFIMdef}), \eqref{eq:CRLB_bis} can be rewritten as 
\begin{equation}
\CRB_{\mathrm{bis}}(\p_n)=a_{1,1}[\Ie^{-1}(\Thetantwo)]_{1,1}+a_{2,2}[\Ie^{-1}(\Thetantwo)]_{2,2}.
\label{eq:CRLBp_bis}
\end{equation}
As for the monostatic system in Section~\ref{sec:monoCRLB}, by computing the inverse of $\Ie(\Thetantwo)$ in \eqref{eq:EFIMdef} and replacing the terms on its main diagonal in \eqref{eq:CRLBp_bis}, we obtain the expression in \eqref{eq:CRLBp_bis_exp} for the position estimation \ac{CRLB}. The \ac{PEB} for a bistatic system is then computed according to \eqref{eq:PEB}.

\section{Network of Cooperative Monostatic Sensors}\label{sec:netmonoCRLB}
In this section, we consider an \ac{ISAC} network consisting of $\Nbs$ cooperating monostatic \acp{BS}, tasked with monitoring a given area \cite{FavMatPucPaoXuGio:J25,PucBacGio:J25}. A schematic representation of the considered scenario is shown in Fig.~\ref{fig:scenarios_combined}a. The notation introduced in Section~\ref{sec:monoCRLB} is adopted to represent both the positions of the target and the \acp{BS}, as well as the target parameters associated with each \ac{BS} $n$. Specifically, the global target position is denoted by $\p = [x, y]^\transp$, whereas the position of \ac{BS} $n$ is denoted by $\mathbf{s}_n = [\xbs, \ybs]^\transp$. Moreover, the target parameters observed by each \ac{BS} are indexed by the subscript $n$.

Unlike the single monostatic configuration, where target velocity estimation is not feasible \cite{richards_book}, the cooperative network enables the estimation of both the target position and velocity. Accordingly, this section investigates the estimation accuracy for both position and velocity. Therefore, in addition to the \ac{PEB}, we introduce the \ac{VEB} as the square root of the \ac{CRLB} of the target velocity estimate.

The target is assumed to move with velocity $\mathbf{v} = [v_x, v_y]^\transp$, where $v_x$ and $v_y$ denote the velocity components along the $x$- and $y$-axes, respectively. The complete target state vector is thus defined as $\mathbf{z} = [\mathbf{p}^\transp, \mathbf{v}^\transp]^\transp$.

\subsection{Cooperative Monostatic PEB}\label{sec:PEB_coop}

To compute the \ac{PEB} for a network of cooperative monostatic sensors, we first derive the \ac{EFIM} for cooperative position estimation, denoted by $\Ie(\p)$.
Assuming that each of the 
$\Nbs$ \acp{BS} independently estimates the target position, the global \ac{EFIM} can be expressed as the sum of the locally computed \acp{EFIM}, transformed to the global coordinate frame, as follows
\begin{equation}
\label{eq:coopFIM}
\Ie(\p) = \sum_{n=1}^{\Nbs} \Jn^\transp \Ie(\p_n) \Jn.
\end{equation}
Here, $\Ie(\p_n)$ denotes the \ac{EFIM} of the target position in the local coordinate frame of the $n$th \ac{BS}. The elements of $\Ie(\p_n)$, derived from \eqref{eq:EFIM_p_mono}, are given by
\begin{align}\label{eq:FIM_n}
  [\Ie(\p_n)]_{1,1}&=\xi_n \left[16 \Df^2 (K^2-1) x_n^2 r_n^2+ \frac{c^2 (\Nr^2-1) y_n^2} {\sec^2(\thetarn)}\right]
\nonumber\\
 [\Ie(\p_n)]_{1,2}&=\xi_n x_n y_n \left[16 \Df^2 (K^2-1) r_n^2-\frac{c^2 (\Nr^2-1)} {\sec^2(\thetarn)}\right]
   \nonumber\\
[\Ie(\p_n)]_{2,1}&=[\Ie(\p_n)]_{1,2}
\nonumber\\
[\Ie(\p_n)]_{2,2}&=\xi_n \left[16 \Df^2 (K^2-1) y_n^2 r_n^2+ \frac{c^2 (\Nr^2-1) x_n^2} {\sec^2(\thetarn)}\right]
\end{align}
with 
\begin{equation}
\xi_n= \frac{\pi^2 K M \Nr \SNR^{(n)}}{6 \, \eta c^2 r_n^4}.
\end{equation}
Since each \ac{BS} operates in its local coordinate system, the Jacobian matrix $\Jn$ accounts for the transformation $n:~\p ~\rightarrow~\p_n$ to the global frame given in \eqref{eq:global_to_local}, and is defined as
\begin{equation}\label{eq:Jn}
\Jn=
\begin{bmatrix}
\frac{\partial x_n}{\partial x} & \frac{\partial x_n}{\partial y} \\
\frac{\partial y_n}{\partial x} & \frac{\partial y_n}{\partial y}
\end{bmatrix}
=
\begin{bmatrix}
\cos(\thetan) & \sin(\thetan)\\
-\sin(\thetan) & \cos(\thetan)
\end{bmatrix}.
\end{equation}

After computing \eqref{eq:coopFIM}, the \ac{CRLB} for position estimation in a cooperative system composed of $\Nbs$ can be numerically obtained as 
\begin{equation}\label{eq:CRLB_coop_mono}
\CRB(\p)=\Tr(\Ie^{-1}(\p)).
\end{equation}
Lastly, the \ac{PEB} is obtained as the square root of \eqref{eq:CRLB_coop_mono} according to \eqref{eq:PEB}.

\subsection{Cooperative Monostatic VEB}\label{sec:VEB_coop}

To derive the \ac{VEB}, we first introduce the following relationships between target local parameters $(f_{\mathrm{D},n},\tau_n,\theta_{\mathrm{R},n})$ and the global parameters $(x, y, v_x, v_y)$, as follows
\begin{align}
    f_{\mathrm{D},n} & = \frac{2}{\lambda_\mathrm{c}} v_{\mathrm{r},n} = \frac{2}{\lambda_\mathrm{c}} \frac{x'_n v_x + y'_n v_y}{\|\mathbf{p} - \mathbf{s}_n\|} \nonumber \\
    \tau_n & = \frac{2}{c} \|\mathbf{p} - \mathbf{s}_n\| \\
    \theta_{\mathrm{R},n} & = \arctan(y'_n/x'_n) - \vartheta_n \nonumber
\end{align}
where $v_{\mathrm{r},n}$denotes the radial component of the target velocity relative to \ac{BS} $n$, while, as recalled from Section~\ref{sec:monoCRLB}, the coordinate differences are defined as $x_n'=x-\xbs$ and $y_n'=y-\ybs$.

We define the Jacobian $\mathbf{J}_\mathrm{\Omega_n}$ of the transformation $\Omega_n: (x, y, v_x, v_y) \rightarrow (f_{\mathrm{D}_n},\tau_n,\theta_{\mathrm{R},n})$ as
\begin{align}\label{eq:Jomega}
& \mathbf{J}_{\Omega_n} =
\begin{bmatrix}
\frac{\partial \fdn}{\partial x} & \frac{\partial \fdn}{\partial y} & \frac{\partial \fdn}{\partial v_x} & \frac{\partial \fdn}{\partial v_y}  \\
\frac{\partial \tau_{n}}{\partial x} & \frac{\partial \tau_{n}}{\partial y} & \frac{\partial \tau_{n}}{\partial v_x} & \frac{\partial \tau_{n}}{\partial v_y} \\
\frac{\partial \thetarn}{\partial x} & \frac{\partial \thetarn}{\partial y} & \frac{\partial \thetarn}{\partial v_x} & \frac{\partial \thetarn}{\partial v_y} \\
\end{bmatrix} \\
& = \begin{bmatrix}
\frac{2}{\lambda_\mathrm{c}}\frac{y'_n(-x'_n\,v_y + y'_n\,v_x)}{r_n^3} & \frac{2}{\lambda_\mathrm{c}}\frac{x'_n(x'_n\,v_y - y'_n\,v_x)}{r_n^3} & \frac{2}{\lambda_\mathrm{c}}\frac{x'_n}{r_n} & \frac{2}{\lambda_\mathrm{c}}\frac{y'_n}{r_n}  \\
\frac{2}{c}\frac{x'_n}{r_n} & \frac{2}{c}\frac{y'_n}{r_n}  & 0 & 0 \\
-\frac{y'_n}{r_n^2} & \frac{x'_n}{r_n^2} & 0 & 0 \\
\end{bmatrix}. \nonumber
\end{align} 

Now, we seek to derive the \ac{EFIM} of joint velocity and position estimation. Similar to the procedure illustrated in Section~\ref{sec:monoCRLB}, for a given \ac{BS} $n$, we partition the parameter vector ${\boldsymbol{\Theta}_n}=[\alpha_n,\phi_n,\fdn,\tau_n,\thetarn]^\transp$ into ${\boldsymbol{\Theta}_n}=[\boldsymbol{\Theta}_{n,1}^\transp,\boldsymbol{\Theta}_{n,2}^\transp]^\transp$ with $\boldsymbol{\Theta}_{n,1}=[\alpha_n,\phi_n]^\transp$ $\boldsymbol{\Theta}_{n,2}=[f_{\mathrm{D},n},\tau_n,\theta_{\mathrm{R},n}]^\transp$
which induces the partition of the \ac{FIM} as \eqref{eq:FIMpart},
%
%
where $\mathbf{A}\in \mathbb{R}^{2\times 2}$, $\mathbf{B}\in \mathbb{R}^{2\times 3}$, and $\mathbf{C}\in \mathbb{R}^{3\times 3}$.
By computed the Schur complement of $\mathbf{A}$ we get the \ac{EFIM}, as follows
\begin{align}\label{eq:EFIM_Doppler}
&\Ie(\boldsymbol{\Theta}_{n,2}) = \mathbf{C}-\mathbf{B}^\transp\mathbf{A}^{-1}\mathbf{B} = \SNR^{(n)}\Nr K M /\eta \\
& \qquad \times
\begin{bmatrix}
\frac{2\pi^2\Ts^2 (M^2-1)}{3} & 0 & 0 \\
0 & \frac{2\pi^2\Df^2 (K^2-1)}{3} & 0 \\
0 & 0 & \frac{\pi^2(\Nr^2-1)cos^2(\thetarn)}{6}\nonumber
\end{bmatrix}. 
\end{align}
Next, we perform reparameterization to calculate the \ac{EFIM} for $(x,y,v_x,v_y)$. This can be obtained by the transformation 
\begin{align}\nonumber
\Ie^{(n)}(x,y,v_x,v_y)& =\mathbf{J}_{\Omega_{n}}^{\transp} \Ie(\Thetantwo)\mathbf{J}_{\Omega_{n}} \\
& =\frac{\SNR^{(n)}\Nr K M}{\eta}
\begin{bmatrix}
   \boldsymbol{\mathcal{I}}^{(n)}_\mathbf{p} &  \boldsymbol{\mathcal{I}}^{(n)}_\mathbf{pv}\\
   \boldsymbol{\mathcal{I}}_\mathbf{pv}^{(n)\transp} & \boldsymbol{\mathcal{I}}^{(n)}_\mathbf{v}
\end{bmatrix}
\label{eq:EFIM_xyv}
\end{align}
where $\mathbf{J}_{\Omega_{n}}$ is given in \eqref{eq:Jomega}, $\boldsymbol{\mathcal{I}}^{(n)}_\mathbf{p} \in \mathbb{R}^{2 \times 2}$ retains information about $\mathbf{p}$, $\boldsymbol{\mathcal{I}}^{(n)}_\mathbf{v} \in \mathbb{R}^{2 \times 2}$ captures information about $\mathbf{v}$, and $\boldsymbol{\mathcal{I}}^{(n)}_\mathbf{pv} \in \mathbb{R}^{2 \times 2}$ represents the position-velocity cross-information. 

Treating the position $\mathbf{p}_n$ as nuisance parameters, the local \ac{EFIM} for the velocity components $(v_x, v_y)$ associated with \ac{BS} $n$, is obtained by computing the Schur complement of $\boldsymbol{\mathcal{I}}^{(n)}_\mathbf{p}$
\begin{align}\label{eq:EFIM_vxvy}
\Ie^{(n)}(v_x,v_y) & = \boldsymbol{\mathcal{I}}^{(n)}_\mathbf{v}-\boldsymbol{\mathcal{I}}_\mathbf{pv}^{(n)\transp}\left(\boldsymbol{\mathcal{I}}_\mathbf{p}^{(n)}\right)^{-1}\boldsymbol{\mathcal{I}}^{(n)}_\mathbf{pv} \nonumber \\
& = k_n
\begin{bmatrix}
x'^{2}_n & x'_n y'_n \\
x'_n y'_n & y'^{2}_n
\end{bmatrix}
\end{align}
where 
\begin{equation}
k_n = \tfrac{\, 8\, \SNR^{(n)} \Nr K M \pi^2 \Ts^2 (M^2-1) (\Nr^2-1) \cos^2(\thetarn) }{48 \eta(M^2-1)\Ts^2 (x'_n v_y - y'_n v_x)^2 + 3(\Nr^2-1) r^2_n \lambda_\mathrm{c}^2 \cos^2(\thetarn)}.
\end{equation}
%

Since each \ac{BS} can only measure the radial component of velocity $v_{\mathrm{r},n}$, the individual \ac{EFIM} $\Ie^{(n)}(v_x,v_y)$ is rank-deficient (i.e., rank one), and thus not invertible. However, by leveraging multiple monostatic sensors, the ambiguity in the tangential component can be resolved.

Assuming that the observations from different \acp{BS} are conditionally independent given the target state $\mathbf{z}$, the global \ac{EFIM} for cooperative velocity estimation can be obtained by summing the local \acp{EFIM}, as follows
\begin{equation}\label{eq:coopFIMvel}
    \Ie(\mathbf{v}) = \sum_{n=1}^{N_\mathrm{BS}} \Ie^{(n)}(v_x,v_y) =
    \begin{bmatrix}
    V_{xx} & V_{xy} \\
    V_{xy} & V_{yy}
    \end{bmatrix}
\end{equation}
where, from \eqref{eq:EFIM_vxvy}, we define the aggregated components as 
\begin{equation}
V_{xx} = \sum_{n=1}^{\Nbs} k_n x'^{2}_n,\quad V_{xy} = \sum_{n=1}^{\Nbs} k_n x'_n y'_n, \quad V_{yy} = \sum_{n=1}^{\Nbs} k_n y'^{2}_n.
\end{equation}

Reparameterization can then be performed to obtain \ac{EFIM} for $(|\mathbf{v}|, \angle{\mathbf{v}})$, as follows
\begin{equation}
\Ie(|\mathbf{v}|,\angle{\mathbf{v}}) =\mathbf{J}_\Gamma^\transp \Ie(\mathbf{v})\mathbf{J}_\Gamma.
\label{eq:EFIM_mod_phase_v}
\end{equation}
In \eqref{eq:EFIM_mod_phase_v}, $\mathbf{J}_\Gamma$ is the Jacobian of the transformation $\Gamma: \{v_x = |\mathbf{v}|\cos({\angle{\mathbf{v}}}),v_y = |\mathbf{v}|\sin({\angle{\mathbf{v}}})\}$, given by
\begin{equation} \label{eq:jac}
    \mathbf{J}_\Gamma = \frac{\partial (v_x, v_y)}{\partial (|\mathbf{v}|, \angle{\mathbf{v}})} =  
    \begin{bmatrix}
    \cos({\angle{\mathbf{v}}}) & -|\mathbf{v}|\sin({\angle{\mathbf{v}}}) \\
    \sin({\angle{\mathbf{v}}}) & |\mathbf{v}|\cos({\angle{\mathbf{v}}})
    \end{bmatrix}.
\end{equation}
Then, by inverting \eqref{eq:EFIMv} at the top of the next page, we obtain the expression of the \ac{CRLB} for the parameters $(|\mathbf{v}|, \angle{\mathbf{v}})$, i.e.,
\begin{figure*}[!t]
\setcounter{MYtempeqncnt}{\value{equation}}
\setlength{\arraycolsep}{2pt}
\begin{align}\label{eq:EFIMv}
&\Ie(|\mathbf{v}|,\angle{\mathbf{v}})= \\
& = \begin{bmatrix}
S_{xx} \cos^2(\angle{\mathbf{v}}) + S_{yy} \sin^2(\angle{\mathbf{v}}) + S_{xy} \sin(2\angle{\vv}) & \quad
S_{xy} |\mathbf{v}| \cos(2\angle{\mathbf{v}}) + (-S_{xx} + S_{yy}) |\mathbf{v}| \cos(\angle{\mathbf{v}}) \sin(\angle{\vv}) \\
S_{xy} |\vv| \cos(2\angle{\vv}) + (-S_{xx} + S_{yy}) |\vv| \cos(\angle{\vv}) \sin(\angle{\vv}) & \quad
|\vv|^2 \Bigl(S_{yy} \cos^2(\angle{\vv}) - 2 S_{xy} \cos(\angle{\vv}) \sin(\angle{\vv}) + S_{xx} \sin^2(\angle{\vv}) \Bigr) \nonumber
\end{bmatrix}
\end{align}
\hrulefill
\vspace*{4pt}
\end{figure*}
\begin{align}\label{eq:CRLB_v}  
\CRB(&|\mathbf{v}|)=\left[\Ie^{-1}(|\mathbf{v}|,\angle{\mathbf{v}})\right]_{1,1} \\ \nonumber & =
  \frac{V_{yy} \cos^2(\angle{\mathbf{v}})  + V_{xx} \sin^2(\angle{\mathbf{v}}) - V_{xy} \sin(2\angle{\mathbf{v}})}{-V_{xy}^2 + V_{xx} V_{yy}} \\ \nonumber \\ 
   \CRB(&\angle{\mathbf{v}})=\left[\Ie^{-1}(|\mathbf{v}|,\angle{\mathbf{v}})\right]_{2,2} \\ \nonumber & =
    \frac{V_{xx} \cos^2(\angle{\mathbf{v}}) + V_{yy} \sin^2(\angle{\mathbf{v}}) + V_{xy} \sin(2\angle{\mathbf{v}})}{\left(-V_{xy}^2 + V_{xx} V_{yy} \right) |\mathbf{v}|^2}.
    \\ \nonumber 
\end{align}
The \ac{VEB} is then defined as 
\begin{equation}
    \mathrm{VEB} = \sqrt{\CRB(|\mathbf{v}|)}.
\label{eq:VEB}
\end{equation}

\emph{Remark:} The bound on the velocity estimation error should, in principle, be computed by first summing the \ac{EFIM} of $(x, y, v_x, v_y)$ in \eqref{eq:EFIM_xyv} across all \acp{BS}, and then extracting $\Ie(\mathbf{v})$ via the Schur complement, as shown in \eqref{eq:EFIM_vxvy}, following an approach analogous to the derivation of the \ac{PEB}. This method yields the tightest possible bound for any unbiased estimator but does not admit a closed-form expression. To balance tractability and accuracy, we adopt an analytical approximation, which was found through numerical simulations to yield results in close agreement with the exact \ac{VEB} across a wide range of practical system parameters. In particular, the approximation becomes especially accurate when the number of receiving antennas is sufficiently large. This suggests that the proposed closed-form expression offers a reliable and insightful performance bound for practical system design.

\section{Multistatic Network}\label{sec:netmultiCRLB}

In this section, we consider an \ac{ISAC} multistatic network consisting of \(N_\mathrm{bis}\) bistatic pairs that cooperate to monitor a given area \cite{DehPucJunGioPaoCai:J24}. A schematic representation is given in Fig.~\ref{fig:scenarios_combined}b. Specifically, the network comprises a single \ac{Tx} and \(N_\mathrm{bis}\) \acp{Rx}. To describe the system geometry and target-related parameters for the \ac{Tx}, the \(n\)th \ac{Rx}, and the target, we adopt the same notation as introduced in Section~\ref{sec:bis CRLB}. In particular, the global positions of the \ac{Tx}, \ac{Rx}~\(n\), and the target are denoted by \(\mathbf{s}_\mathrm{T} = [x_\mathrm{T}, y_\mathrm{T}]^\transp\), \(\mathbf{s}_n = [x^{(n)}_\mathrm{R}, y^{(n)}_\mathrm{R}]^\transp\), and \(\mathbf{p} = [x, y]^\transp\), respectively.

Following the approach adopted in the monostatic network case (Section~\ref{sec:netmonoCRLB}), the analysis is extended to account for target velocity. In particular, whereas a single bistatic pair is insufficient for velocity estimation, cooperation among multiple pairs in a multistatic configuration enables such estimation \cite{willis05}. We assume the target moves with velocity \(\mathbf{v} = [v_x, v_y]^\transp\), where \(v_x\) and \(v_y\) denote its components along the \(x\)- and \(y\)-axes. The full state of the target is then given by \(\mathbf{z} = [\mathbf{p}^\transp, \mathbf{v}^\transp]^\transp\).


\subsection{Cooperative Multistatic PEB}\label{sec:PEB_multi}
Analogously to the network of monostatic sensors described in Section~\ref{sec:netmonoCRLB}, we assume that each of the $N_\mathrm{bis}$ bistatic pairs independently estimates the target position.

Accordingly, the global \ac{EFIM} $\Ie(\p)$ for the target position is formulated as in \eqref{eq:coopFIM}. In this setting, $\Ie(\p_n)$ represents the \ac{EFIM} of the target position expressed in the local coordinate frame of the $n$th \ac{Rx}, with its elements computed according to \eqref{eq:EFIM_p_bis}.

Once $\Ie(\p)$ is numerically evaluated, the \ac{PEB} for a multistatic configuration—comprising a single \ac{Tx} and multiple \acp{Rx}—is obtained as the square root of the expression in \eqref{eq:CRLB_coop_mono}.

\subsection{Cooperative Multistatic VEB}
\label{sec:VEB_multi}
For a bistatic radar setup, we have the following relationships between the target local parameters $(f_{\mathrm{D},n},\tau_n,\theta_{\mathrm{R},n})$ and the global parameters $(x, y, v_x, v_y)$ \cite{willis05}
\begin{align}
    f_{\mathrm{D},n} & = \frac{1}{\lambda_\mathrm{c}} \frac{\mathbf{v}^{\mathrm{T}}(\mathbf{p}-\mathbf{s}_\mathrm{T})}{\|\mathbf{p} - \mathbf{s}_\mathrm{T}\|} + \frac{1}{\lambda_\mathrm{c}} \frac{\mathbf{v}^{\mathrm{T}}(\mathbf{p}-\mathbf{s}_n)}{\|\mathbf{p} - \mathbf{s}_n\|} \nonumber \\
    \tau_n & = \frac{1}{c} \|\mathbf{p} - \mathbf{s}_\mathrm{T}\| + \frac{1}{c} \|\mathbf{p} - \mathbf{s}_n\| \\
    \theta_{\mathrm{R},n} & = \arctan(y'_n/x'_n) - \vartheta_n \nonumber.
\end{align}

\begin{figure*}[!t]
\setlength{\arraycolsep}{2pt}
{\small
\begin{align}
    &\mathbf{J}_{\Omega_n} = \label{eq:JomegaBis}\\
    &=\begin{bmatrix}
    \frac{v_x}{\lambda_\mathrm{c}}\left(\frac{y'^{2}_\mathrm{T}}{r_\mathrm{T}^3} + \frac{y'^{2}_n}{r_n^3}\right) - \frac{v_y}{\lambda_\mathrm{c}} \left(\frac{x'_\mathrm{T}\,y'_\mathrm{T}}{r_\mathrm{T}^3} + \frac{x'_n\,y'_n}{r_n^3}\right) & 
    -\frac{v_x}{\lambda_\mathrm{c}}\left(\frac{x'_\mathrm{T}\,y'_\mathrm{T}}{r_\mathrm{T}^3} + \frac{x'_n\,y'_n}{r_n^3}\right) + \frac{v_y}{\lambda_\mathrm{c}}\left(\frac{x'^{2}_\mathrm{T}}{r_\mathrm{T}^3} + \frac{x'^{2}_n}{r_n^3}\right) & 
    \frac{x'_\mathrm{T}}{\lambda_\mathrm{c}\,r_\mathrm{T}} + \frac{x'_n}{\lambda_\mathrm{c}\,r_n} & \frac{y'_\mathrm{T}}{\lambda_\mathrm{c}\,r_\mathrm{T}} + \frac{y'_n}{\lambda_\mathrm{c}\,r_n}  \\
    \frac{x'_\mathrm{T}}{c\,r_\mathrm{T}} + \frac{x'_n}{c\,r_n} &  \frac{y'_\mathrm{T}}{c\,r_\mathrm{T}} + \frac{y'_n}{c\,r_n}  & 0 & 0 \\
    -\frac{y'_n}{r_n^2} & \frac{x'_n}{r_n^2} & 0 & 0 \\
    \end{bmatrix} \nonumber \\
    &x'_\mathrm{T}=x-x_\mathrm{T},\, y'_\mathrm{T}=y-y_\mathrm{T},\, r_n = \|\mathbf{p}-\mathbf{s}_n\|,\, r_\mathrm{T} = \|\mathbf{p}-\mathbf{s}_\mathrm{T}\|
\end{align}
}
\hrulefill
\vspace*{4pt}
\end{figure*}

Considering a given \ac{Rx} $n$, we partition the parameter vector ${\boldsymbol{\Theta}_n}=[\alpha_n,\phi_n,f_{\mathrm{D},n},\tau_n,\theta_{\mathrm{R},n}]^\transp$ into ${\boldsymbol{\Theta}_n}=[\boldsymbol{\Theta}_{n,1}^\transp,\boldsymbol{\Theta}_{n,2}^\transp]^\transp$ with $\boldsymbol{\Theta}_{n,1}=[\alpha_n,\phi_n]^\transp$ $\boldsymbol{\Theta}_{n,2}=[f_{\mathrm{D},n},\tau_n,\theta_{\mathrm{R},n}]^\transp$. Consequently, we get the same \ac{EFIM} $\Ie(\Thetantwo)$ as in \eqref{eq:EFIM_Doppler}. \\
Next, we perform a parameterization to compute the \ac{EFIM} for the state vector $(x, y, v_x, v_y)$, as defined in \eqref{eq:EFIM_xyv}. The matrix $\mathbf{J}_{\Omega_n}$, which appears in that expression, is defined in \eqref{eq:Jomega} and its bistatic counterpart is provided in \eqref{eq:JomegaBis} at the top of the next page.

The \ac{EFIM} of $(v_x, v_y)$ is given in \eqref{eq:EFIM_vxvyBIS} at the top of the next page. This is obtained by treating $(x,y)$ as nuisance parameters and computing the Schur complement of $\boldsymbol{\mathcal{I}}^{(n)}_\mathbf{p}$. 

\emph{Remark:}
It is worth noting that, under the assumptions $r_\mathrm{T} = r_n$, $x'_\mathrm{T} = x'_n$, and $y'_\mathrm{T} = y'_n$, the bistatic configuration degenerates into the monostatic case. Consequently, the matrix in \eqref{eq:EFIM_vxvyBIS} simplifies to that in \eqref{eq:EFIM_vxvy}.

The \ac{EFIM} for cooperative target velocity estimation is given by
\begin{equation}\label{eq:coopFIMvelBIS}
    \Ie^{\text{multi}}(\mathbf{v}) = \sum_{n=1}^{N_\mathrm{bis}} \Ie^{(n)}(v_x,v_y) =
    \begin{bmatrix}
    V_{xx}^{\text{multi}} & V_{xy}^{\text{multi}} \\
    V_{xy}^{\text{multi}} & V_{yy}^{\text{multi}}
    \end{bmatrix}.
\end{equation}
where we define $V_{xx}^{\text{multi}} = \sum_{n=1}^{N_\mathrm{bis}} k_n (r_\mathrm{T} x_n' + r_n x_\mathrm{T}')^2$, $V_{xy}^{\text{multi}} = \sum_{n=1}^{N_\mathrm{bis}} k_n (r_\mathrm{T} x_n' + r_n x_\mathrm{T}')(r_\mathrm{T} y_n' + r_n y_\mathrm{T}')$, and $V_{yy}^{\text{multi}} = \sum_{n=1}^{N_\mathrm{bis}} k_n (r_\mathrm{T} y_n' + r_n y_\mathrm{T}')^2$.

From \eqref{eq:coopFIMvelBIS} and the Jacobian $\mathbf{J}_\Gamma$ in \eqref{eq:jac}, it is possible to obtain the \ac{EFIM} for $(|\mathbf{v}|, \angle{\mathbf{v}})$ by computing 
$\Ie(|\mathbf{v}|,\angle{\mathbf{v}})$ as in \eqref{eq:EFIM_mod_phase_v}.
By inverting this matrix, we obtain the expression of the \ac{CRLB} for each parameter as
\begin{align}
&\CRB(|\mathbf{v}|)=\left[\Ie^{-1}(|\mathbf{v}|,\angle{\mathbf{v}})\right]_{1,1} \nonumber \\  
&= \frac{V_{yy}^{\text{multi}} \cos^2(\angle{\mathbf{v}})  + V_{xx}^{\text{multi}} \sin^2(\angle{\mathbf{v}}) - V_{xy}^{\text{multi}} \sin(2\angle{\mathbf{v}})}{-(V_{xy}^{\text{multi}})^2 + V_{xx}^{\text{multi}} V_{yy}^{\text{multi}}} \label{eq:CRLBbis_v} 
\end{align}
\begin{align}  &\CRB(\angle{\mathbf{v}})=\left[\Ie^{-1}(|\mathbf{v}|,\angle{\mathbf{v}})\right]_{2,2} \nonumber \\ 
& = \frac{V_{xx}^{\text{multi}} \cos^2(\angle{\mathbf{v}}) + V_{yy}^{\text{multi}} \sin^2(\angle{\mathbf{v}}) + V_{xy}^{\text{multi}} \sin(2\angle{\mathbf{v}})}{\left(-(V_{xy}^{\text{multi}})^2 + V_{xx}^{\text{multi}} V_{yy}^{\text{multi}} \right) |\mathbf{v}|^2} 
\end{align}
from which the \ac{VEB} can be calculated as in \eqref{eq:VEB}.

\begin{figure*}[!t]
\setlength{\arraycolsep}{2pt}
\begin{align}
    \Ie^{(n)}(v_x,v_y) =& \boldsymbol{\mathcal{I}}_\mathbf{v}-\boldsymbol{\mathcal{I}}_\mathbf{pv}^\transp\boldsymbol{\mathcal{I}}_\mathbf{p}^{-1}\boldsymbol{\mathcal{I}}_\mathbf{pv} = k_n
    \begin{bmatrix}
        (r_\mathrm{T} x_n' + r_n x_\mathrm{T}')^2 & (r_\mathrm{T} x_n' + r_n x_\mathrm{T}')(r_\mathrm{T} y_n' + r_n y_\mathrm{T}') \\
        (r_\mathrm{T} x_n' + r_n x_\mathrm{T}')(r_\mathrm{T} y_n' + r_n y_\mathrm{T}') & (r_\mathrm{T} y_n' + r_n y_\mathrm{T}')^2
    \end{bmatrix} \label{eq:EFIM_vxvyBIS} \\
    k_n =& \frac{A r_\mathrm{T}^4\cos^2{\theta_{\mathrm{R},n}} (r_\mathrm{T}(x'^2_n+y'^2_n)+r_n(x'_nx'_\mathrm{T}+y'_ny'_\mathrm{T}))^2}{12\Delta f^2\Ts^2(K^2-1)(M^2-1)\Upsilon_1^2 + 3(\Nr^2-1)r_n^2r_\mathrm{T}^2\cos^2{\theta_{\mathrm{R},n}}\Upsilon_2} \\
    A =& 2\pi^2\Delta f^2 \Ts^2 \SNR^{(n)}K(K^2-1)M(M^2-1)\Nr(\Nr^2-1)/\eta \\
    \Upsilon_1 =& r_\mathrm{T}^4(v_yx'_n-v_xy'_n)(x'^2_n+y'^2_n) + r_nr_\mathrm{T}^3(v_yx'_n-v_xy'_n)(x'_nx'_\mathrm{T}+y'_ny'_\mathrm{T}) \nonumber \\
    &+r_n^3r_\mathrm{T}(v_yx'_\mathrm{T}-v_xy'_\mathrm{T})(x'_nx'_\mathrm{T}+y'_ny'_\mathrm{T}) + r_n^4(v_yx'_\mathrm{T}-v_xy'_\mathrm{T})(x'^2_\mathrm{T}+y'^2_\mathrm{T}) \\
    \Upsilon_2 =& c^2\Ts^2(M^2-1)r_n^2(v_y x'_\mathrm{T}-v_xy'_\mathrm{T})^2(x'_\mathrm{T}y'_n-x'_ny'_\mathrm{T})^2 \nonumber \\
    &+ \lambda_\mathrm{c}^2\Delta f^2(K^2-1)r_\mathrm{T}^4(r_\mathrm{T}x'^2_n+r_nx'_nx'_\mathrm{T}+r_\mathrm{T} y'^2_n+r_ny'_ny'_\mathrm{T})^2
\end{align}
\hrulefill
\vspace*{4pt}
\end{figure*}



\section{PEB and VEB for a Heterogeneous ISAC Network}\label{sec:netmixCRLB}
For completeness, we also consider heterogeneous \ac{ISAC} networks in which each \ac{BS} may operate either as a monostatic sensor or as the \ac{Tx}/\ac{Rx} node of a bistatic pair. This general setting encompasses both the cooperative monostatic and multistatic configurations introduced in Section~\ref{sec:netmonoCRLB} and Section~\ref{sec:netmultiCRLB}, respectively. 

To derive the \ac{PEB} in such networks, we follow the approaches outlined in Section~\ref{sec:PEB_coop} and Section~\ref{sec:PEB_multi}. Specifically, when performing the summation in \eqref{eq:coopFIM}, the matrix $\Ie(\p_n)$ denotes the \ac{EFIM} of the target position associated with the $n$th node, which can be either (i) the \ac{EFIM} of a monostatic \ac{BS}, as given in \eqref{eq:FIM_n}, or (ii) the \ac{EFIM} of a bistatic \ac{Rx}, as defined in \eqref{eq:EFIM_p_bis}.

Similarly, the \ac{VEB} for a heterogeneous \ac{ISAC} network is obtained by extending the methodology described in Section~\ref{sec:VEB_coop} and Section~\ref{sec:VEB_multi}. In this case, the matrix $\Ie^{(n)}(v_x, v_y)$ in \eqref{eq:coopFIMvel} denotes the \ac{EFIM} for the velocity components $(v_x, v_y)$, corresponding either to a monostatic \ac{BS}, computed as in \eqref{eq:EFIM_vxvy}, or to a bistatic \ac{Rx}, as given in \eqref{eq:EFIM_vxvyBIS}.

\section{Numerical Results}\label{sec:numres}
This section presents numerical results to evaluate the fundamental limits of target localization and velocity estimation accuracy in cooperative monostatic and multistatic \ac{ISAC} network architectures. The bounds derived in the previous sections are employed to analyze the \ac{PEB} and \ac{VEB} as functions of the number of \acp{BS} $\Nbs$ and the number of bistatic pairs $N_\mathrm{bis}$. 

Without loss of generality, the \acp{BS} are placed along the perimeter of a square area of side length $84\,$m,\footnote{This dimension is selected based on the \ac{CP} duration to prevent \ac{ISI} in the monostatic \acp{BS} as well as in the multistatic setup.} with each \ac{BS} employing a \ac{Tx}/\ac{Rx} \ac{ULA} oriented toward the center of the square, as depicted in Fig.~\ref{fig:scenarios_combined}. The system parameters are summarized in Table~\ref{tab:sim_param}, and the transmitted symbols $x[k,m]$ are drawn from a \ac{QPSK} constellation, which implies $\eta=1$. To highlight the fundamental aspects of cooperative sensing, all \acp{BS} are assumed to share the same system parameters. This assumption simplifies the exposition, leaving the straightforward extension to more general scenarios to the reader.

Three types of analyses are conducted. First, we vary the target position within the monitored area while fixing the system parameters according to Table~\ref{tab:sim_param}. The resulting \ac{PEB} and \ac{VEB} are visualized through heatmaps to assess spatial variations in estimation performance. For this analysis, the cooperative monostatic configuration includes four \acp{BS} placed at coordinates (in meters) $\mathbf{s}_1 = [42, 0]^\transp$, $\mathbf{s}_2 = [0, 42]^\transp$, $\mathbf{s}_3 = [42, 42]^\transp$, and $\mathbf{s}_4 = [84, 42]^\transp$. The same spatial setup is adopted for the multistatic configuration, where $\mathbf{s}_1$ denotes the \ac{Tx} position, and $\mathbf{s}_2$, $\mathbf{s}_3$, and $\mathbf{s}_4$ correspond to the \acp{Rx}.

Second, we fix the target position and vary key system parameters that influence estimation accuracy. For the \ac{PEB}, we analyze the impact of the bandwidth fraction for sensing $\rho_\mathrm{f}$ (which affects range resolution) and the number of receive antennas $\Nr$ (which influences \ac{AoA} resolution). For the \ac{VEB}, we consider the effect of the number of \ac{OFDM} symbols $\rho_\mathrm{t}$ (which affects Doppler estimation) and again $\Nr$. This analysis is carried out for different numbers of cooperating monostatic \acp{BS} (up to four) and multistatic \acp{Rx} (up to three), using the same network geometry introduced above.

Finally, two illustrative examples demonstrate how the proposed bounds can inform network design decisions. These include selecting the optimal subset of \acp{BS}, or choosing the most suitable \ac{BS} from a given set to serve as the transmitter in a multistatic configuration, based on the target's location.

Given the dependence of both the \ac{VEB} and $\CRB(\angle \mathbf{v})$ on the target’s velocity direction, all results related to these bounds are averaged over $1000$ Monte Carlo iterations. In each iteration, the velocity angle $\angle \mathbf{v}$ is randomly drawn from the interval $[0, 2\pi)$, while the velocity magnitude is fixed according to the value specified in Table~\ref{tab:sim_param}.


\begin{table} [t] 
\centering
 \caption{ISAC Network Parameters} \label{tab:sim_param}
 \resizebox{0.95\columnwidth}{!}{
 \renewcommand{\arraystretch}{1}
 \begin{tabular}{l |l |l}
\toprule
Parameters [5G NR FR2] & Symbol & Value \\
\midrule
Number of \ac{Tx} antennas & $\Nt$ & $16$ \\
Number of \ac{Rx} antennas & $\Nr$ & $16$ \\
OFDM symbols per frame & $M_\mathrm{f}$ & $1120$\\
Total active subcarriers & $K_\mathrm{a}$ & $3168$ \\
Carrier frequency & $f_\mathrm{c}$ & $28$ GHz \\
Subcarrier spacing & $\Delta f$ & $120$ kHz \\
Total OFDM symbol duration & $T_\mathrm{s}$ & $8.92\,\mu$s \\
Fraction of subcarriers for sensing & $\rho_f$ & $0.2$ \\
Fraction of OFDM symbols for sensing & $\rho_t$ & $0.1$ \\
Total OFDM signal power & $P_\mathrm{T}$ & $20$ dBm\\
Power per subcarrier & $P_\mathrm{avg}$ & $-15\,$dBm \\
Noise power spectral density & $N_0$ & $4\cdot 10^{-20}$ W/Hz\\
Target radar cross-section & $\sigma$ & $1$ m$^2$\\
Target speed & $|\mathbf{v}|$ & $22$ m$/$s\\
 \bottomrule
\end{tabular}
}
\label{NRparam}
\end{table}

\subsection{PEB vs Target Position}\label{sec:PEBvsPos}
This subsection analyzes target localization accuracy across the area of interest using heatmaps. The target is moved throughout the area, and for each position, the \ac{PEB} is computed following the formulations in Sections~\ref{sec:netmonoCRLB} and~\ref{sec:netmultiCRLB}. It is worth noting that, for an unbiased estimator, the \ac{PEB} represents a lower bound on the \ac{RMSE} of position estimation. As such, this analysis offers valuable insights into the coverage performance of an \ac{ISAC} network with a given number $\Nbs$ of cooperative monostatic \acp{BS} or $N_\mathrm{bis}$ bistatic pairs. For instance, by setting a threshold on the maximum acceptable \ac{PEB}, one can characterize the spatial extent of effective sensing coverage. 

Referring to the parameters in Table~\ref{tab:sim_param}, we define the sensing transmit power at each active \ac{Tx} as
\begin{equation}
 P_\mathrm{sens} = \rho_f P_\mathrm{T}.   
\end{equation}
Then, the total sensing transmit power across the network is denoted by $P_\mathrm{sens,tot} = N_\mathrm{Tx} P_\mathrm{sens}$, where $N_\mathrm{Tx}$ is the number of active \acp{Tx}.
Since all analyzed multistatic configurations involve a single \ac{Tx}, their total sensing power is $P^{\mathrm{multi}}_\mathrm{sens,tot} = P_\mathrm{sens} $. To enable a fair comparison, we fix this total power budget across all network architectures. Accordingly, in cooperative monostatic scenarios with $\Nbs > 1$ transmitting \acp{BS}, the per-transmitter sensing power is scaled as
\begin{equation}
P^{\mathrm{mono}}_\mathrm{sens} = \frac{P^{\mathrm{multi}}_\mathrm{sens, tot}}{\Nbs} = \frac{P_\mathrm{sens}}{\Nbs}.
\label{eq:power_budget}
\end{equation}
This normalization ensures that all configurations operate under the same overall sensing power budget. 

\begin{figure*}[t]
    \centering
    \subfloat[]{\includegraphics[width=0.38\linewidth]{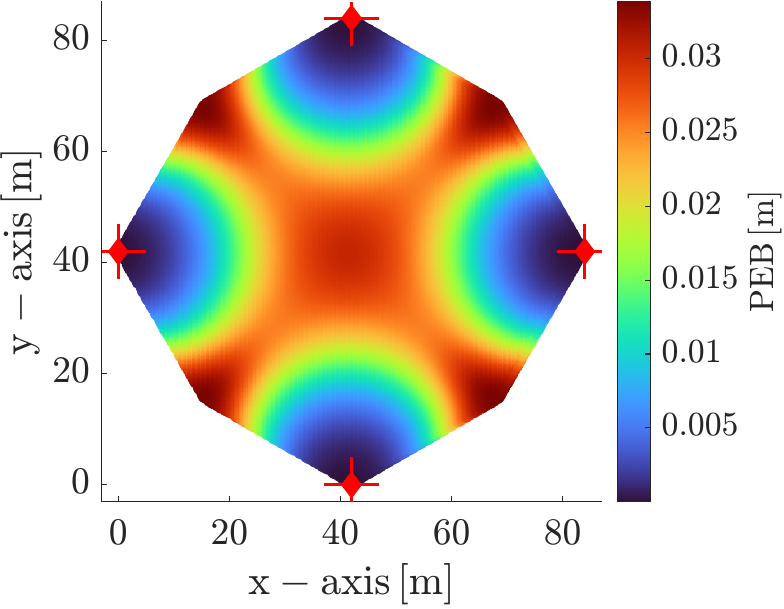}%
    \label{fig:heatmampsMONO}}
    \hspace{1cm}
    \subfloat[]{\includegraphics[width=0.38\linewidth]{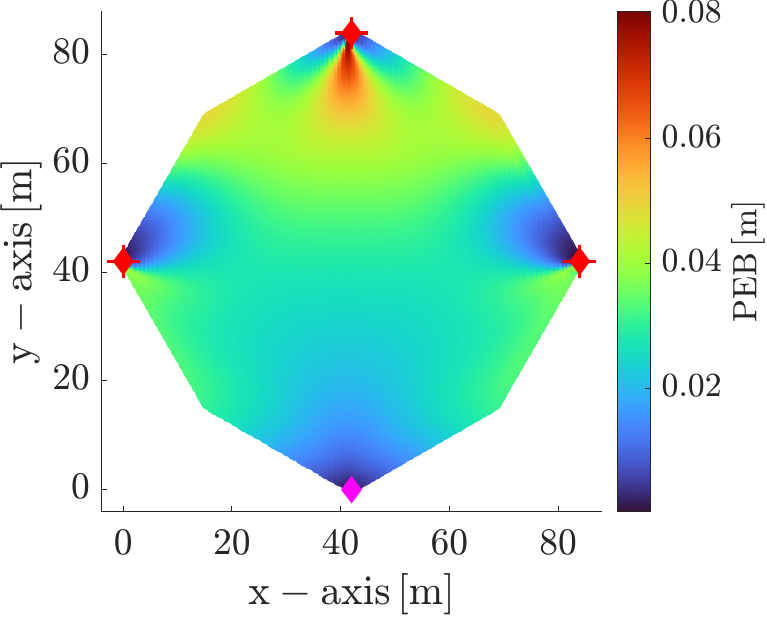}%
    \label{fig:heatmampsMULT}}

    \vspace{1mm}
    
    \subfloat[]{\includegraphics[width=0.38\linewidth]{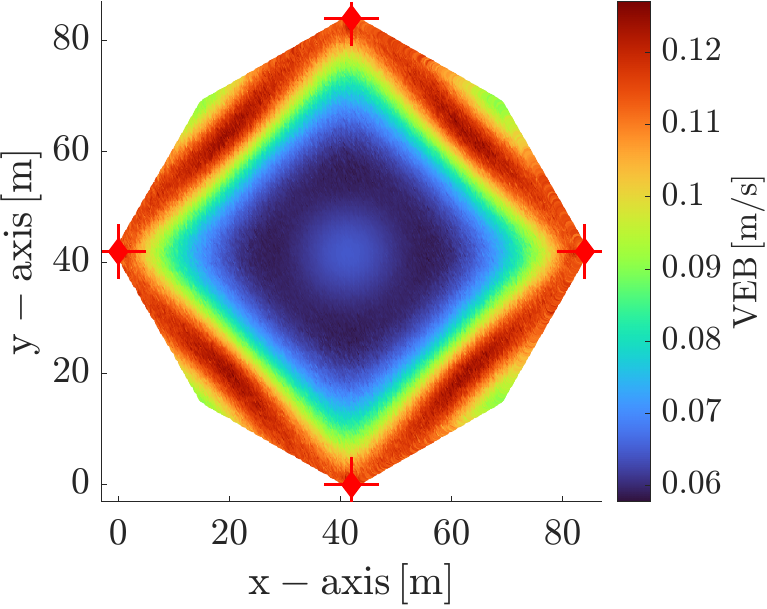}%
    \label{fig:VEBheatmampsMONO}}
    \hspace{1cm}
    \subfloat[]{\includegraphics[width=0.38\linewidth]{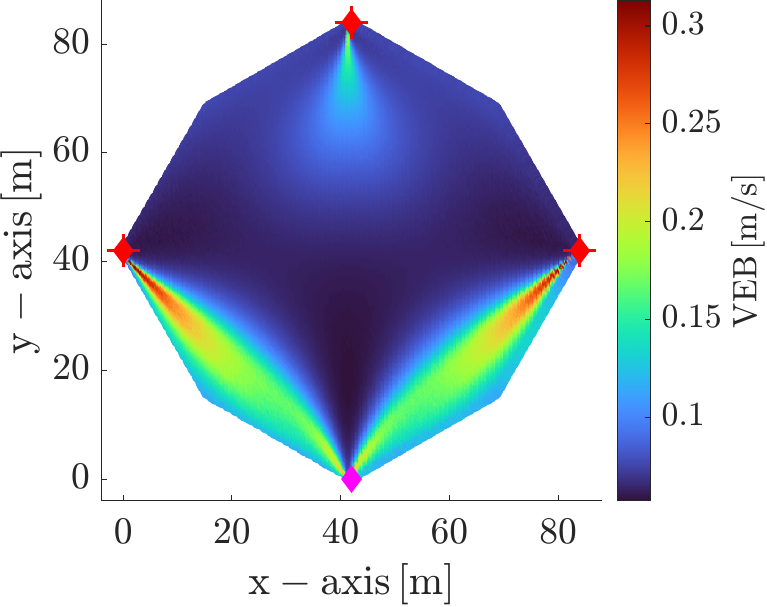}%
    \label{fig:VEBheatmampsMULT}}
    
    \caption{Heatmaps showing: (a) the \ac{PEB} and (c) the \ac{VEB} for an \ac{ISAC} network composed of $\Nbs=4$ monostatic \acp{BS}. (b) the \ac{PEB} and (d) the \ac{VEB} for a multistatic \ac{ISAC} network composed of $N_\mathrm{bis}=3$ bistatic pairs. In (a) and (c), red markers represent the location of the monostatic \acp{BS}; in (b) and (d), the \ac{Rx} locations are indicated by the red marker, while the magenta marker indicates the \ac{Tx} position.}
    \label{fig:hetmap}
\end{figure*}

The corresponding results are shown in Fig.~\ref{fig:heatmampsMONO} and Fig.~\ref{fig:heatmampsMULT}. Fig.~\ref{fig:heatmampsMONO} depicts the \ac{PEB} for a cooperative monostatic network with $\Nbs = 4$ \acp{BS}, while Fig.~\ref{fig:heatmampsMULT} presents the \ac{PEB} for a multistatic configuration comprising $N_\mathrm{bis} = 3$ bistatic pairs with one shared \ac{Tx}.

Notably, for the considered scenario, the cooperative monostatic configuration exhibits superior localization performance, with \ac{PEB} values falling below $0.005\,\mathrm{m}$ in regions near the \acp{BS}. In contrast, the multistatic architecture yields higher \ac{PEB} values, particularly in areas distant from the \ac{Tx} and along the baseline between the \ac{Tx} and the \ac{Rx} located at $[42, 84]\,\mathrm{m}$. For the latter case, this degradation stems from the well-known limitation of bistatic systems in localizing targets that lie on or near the direct path between \ac{Tx} and \ac{Rx} \cite{PucMatPaoGio:C22}.

\subsection{\Ac{VEB} vs Target Position}
The same simulation setup as in Section~\ref{sec:PEBvsPos} is adopted here to analyze the \ac{VEB} across different target positions within the monitored area.

Fig.~\ref{fig:VEBheatmampsMONO} illustrates the \ac{VEB} distribution for a network comprising $\Nbs = 4$ cooperative monostatic \acp{BS}. The highest \ac{VEB} values are observed when the target lies along the line connecting two closely spaced \acp{BS}. In such cases, the \acp{BS} observe nearly identical radial components of the target's velocity, leading to redundant information. As a result, their combined contribution to velocity estimation is comparable to that of a single \ac{BS}, thereby diminishing the benefits of cooperation.

Fig.~\ref{fig:VEBheatmampsMULT} presents the \ac{VEB} distribution for a multistatic scenario composed of $N_\mathrm{bis} = 3$ bistatic pairs, with the transmitter located at $[42, 0]\,$m. Although elevated \ac{VEB} values are observed in some localized regions, the configuration generally enables accurate velocity estimation over a large portion of the monitored area. Nevertheless, its performance is consistently inferior to that of the cooperative monostatic configuration, indicating a lower overall estimation accuracy.

\emph{Remark:}
As will be further discussed, it is particularly noteworthy that the areas with the lowest \ac{PEB} do not necessarily correspond to those with the lowest \ac{VEB}. This observation highlights the need for distinct network design strategies for position and velocity estimation.

\subsection{PEB vs System Parameters}
In this analysis, the target is positioned at fixed coordinates $(70,56)\,\text{m}$, and the \ac{PEB} is evaluated as a function of the fraction $\rho_f$ of subcarriers used for sensing and the number of receive antennas $\Nr$, for both cooperative monostatic and multistatic configurations. Multiple performance curves are generated by varying the number of monostatic \acp{BS}, $\Nbs$, and the number of bistatic pairs, $N_\mathrm{bis}$. The objective is to determine which parameters most significantly impact localization accuracy for the chosen target location.

The same power normalization strategy described in Section~\ref{sec:PEBvsPos} is applied here. In particular, for the monostatic configuration, the total transmit power is scaled according to \eqref{eq:power_budget}, based on the number $\Nbs$ of cooperating \acp{BS}.

\begin{figure}
    \centering
    \includegraphics[width=0.95\columnwidth]{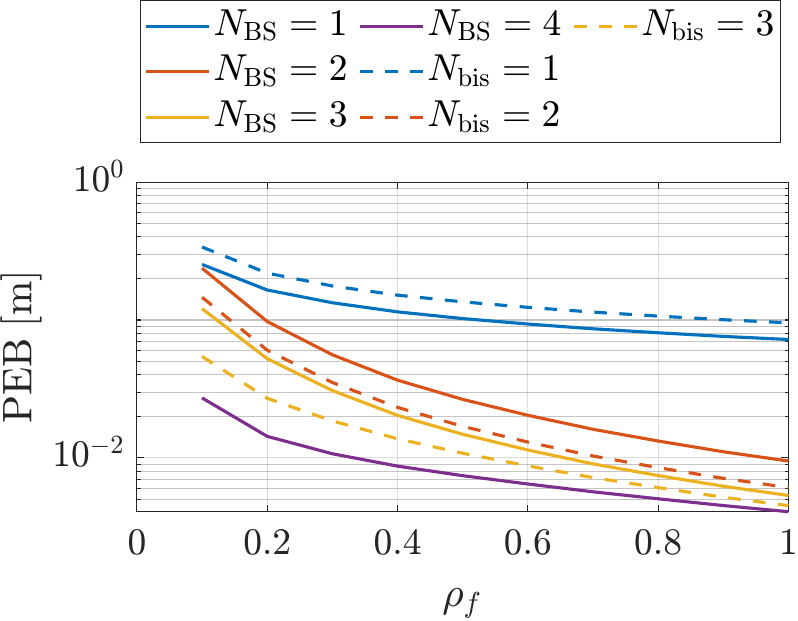}
    \caption{\Ac{PEB} as a function of the fraction $\rho_f$ of subcarriers used for sensing for both cooperative monostatic and multistatic configurations.}
    \label{fig:PEBvsK}
\end{figure}

Fig.~\ref{fig:PEBvsK} shows that the multistatic configuration outperforms its monostatic counterpart when fixing the number of \ac{Tx}-\ac{Rx} pairs for $\Nbs = N_\mathrm{bis} \geq 1$. For example, with a sensing subcarrier fraction of $\rho_f = 0.8$, a cooperative monostatic network with $\Nbs = 2$ \acp{BS} yields a \ac{PEB} greater than $1\,\text{cm}$, whereas a multistatic setup with $N_\mathrm{bis} = 2$ bistatic pairs achieves sub-centimeter accuracy. However, when the total number of \acp{BS} is fixed, regardless of their roles (e.g., $\Nbs = 4$ versus $N_\mathrm{bis} = 3$), the cooperative monostatic network outperforms the multistatic one. 

For both configurations, allocating many subcarriers for sensing may not be advantageous when considering one monostatic \ac{BS} or one bistatic pair. 
Contrariwise, the number of \acp{BS} plays a crucial role in enhancing localization performance. In particular, increasing from $\Nbs = 1$ to $\Nbs = 2$, or equivalently from $N_\mathrm{bis} = 1$ to $N_\mathrm{bis} = 2$, results in a substantial improvement in accuracy.

\begin{figure}
    \centering
    \includegraphics[width=0.95\columnwidth]{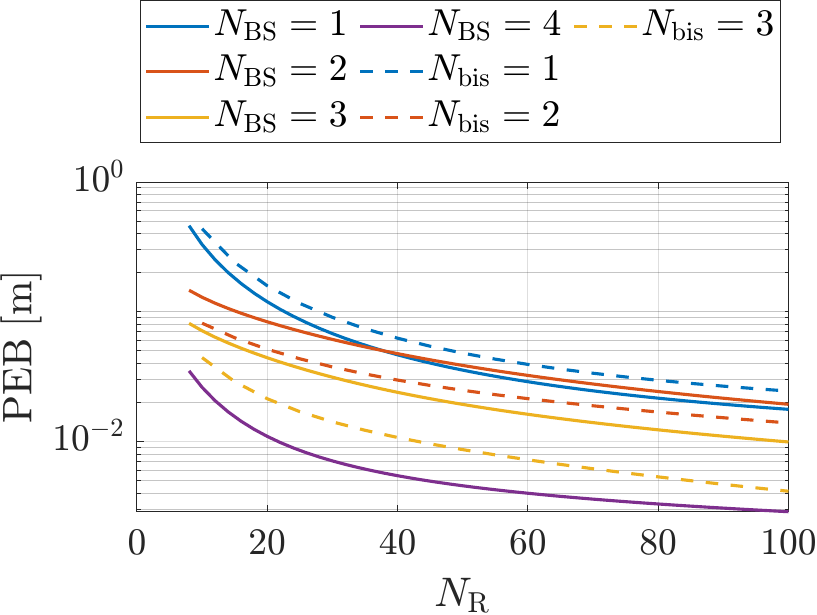}
    \caption{\Ac{PEB} as a function of the number of receive antennas $\Nr$ for both cooperative monostatic and multistatic configurations.}
    \label{fig:PEBvsNr}
\end{figure}

In Fig.~\ref{fig:PEBvsNr}, the impact of the number of antennas at the \ac{Rx}, $\Nr$, is shown by considering up to $100$ antenna elements for both cooperative monostatic and multistatic configurations. Notably, increasing the number of antennas at the \ac{Rx} greatly enhances localization performance. Specifically, sub-centimeter accuracy can be achieved when $\Nr = 100$ and $N_\mathrm{BS} = 4$ or $N_\mathrm{bis} = 3$. The crossing point between the \ac{PEB} curves for one and two monostatic sensors is due to the reduction in transmitted power, as it is divided among the \ac{BS}, and to the fact that the resulting geometry may not be favorable for the specific target location under consideration.

\subsection{\Ac{VEB} vs System Parameters}
In this analysis, the target is positioned at fixed coordinates $(70,30)\,\text{m}$, and the \ac{VEB} is evaluated as a function of the fraction $\rho_t$ of \ac{OFDM} symbols used for sensing and the number of receive antennas $\Nr$, for both cooperative monostatic and multistatic configurations. Multiple performance curves are generated by varying the number of monostatic \acp{BS}, $\Nbs$, and the number of bistatic pairs, $N_\mathrm{bis}$, in each configuration.

To ensure a fair comparison, the same power normalization strategy described in Section~\ref{sec:PEBvsPos} is applied. In particular, the total transmit power across the network is kept constant, and for monostatic configurations, the per-\ac{BS} transmit power is scaled according to \eqref{eq:power_budget}, based on the number $\Nbs$ of transmitting \acp{BS}.

\begin{figure}
    \centering
    \includegraphics[width=0.95\columnwidth]{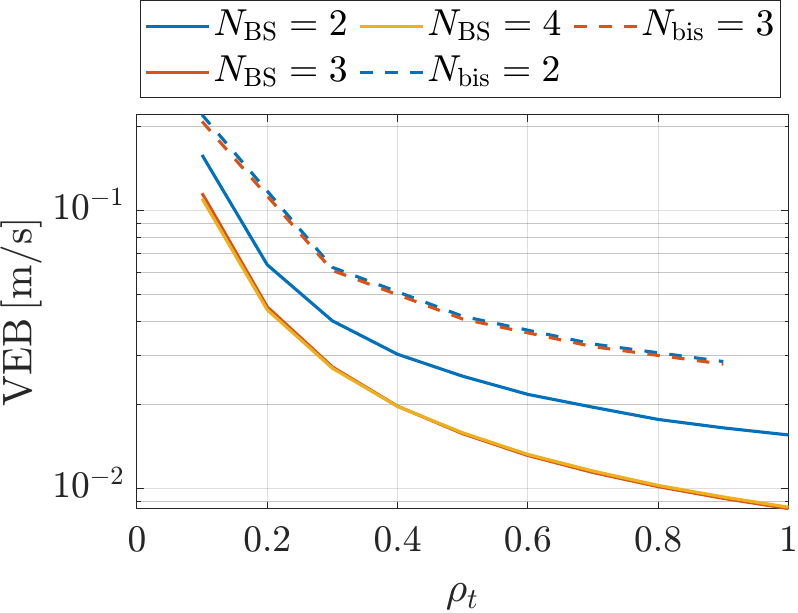}
    \caption{\Ac{VEB} as a function of the fraction $\rho_t$ of \ac{OFDM} symbols used for sensing for both cooperative monostatic and multistatic configurations.}
    \label{fig:VEBvsM}
\end{figure}

\begin{figure}
    \centering
    \includegraphics[width=0.95\columnwidth]{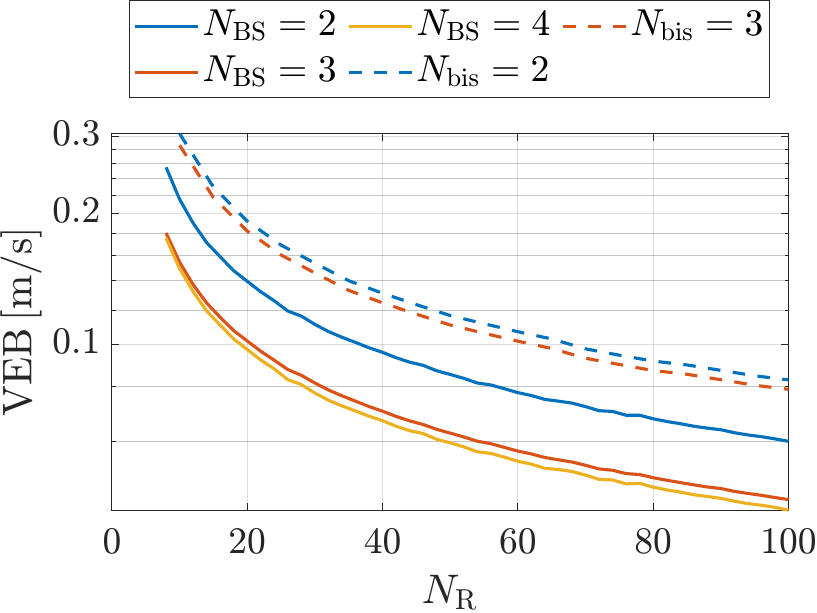}
    \caption{\Ac{VEB} as a function of the number of receive antennas $\Nr$ for both cooperative monostatic and multistatic configurations.}
    \label{fig:VEBvsNr}
\end{figure}

The results are presented in Fig.~\ref{fig:VEBvsM} and Fig.~\ref{fig:VEBvsNr}. For the considered target position, the cooperative monostatic architecture consistently outperforms the multistatic setup in terms of velocity estimation accuracy. This observation aligns with the spatial trends shown in the heatmaps of Fig.~\ref{fig:VEBheatmampsMONO} and Fig.~\ref{fig:VEBheatmampsMULT}. Moreover, the performance of both configurations tends to saturate with increasing numbers of \acp{BS}. The curves for $\Nbs = 3$ and $\Nbs = 4$ are nearly overlapping, as are those for $N_\mathrm{bis} = 2$ and $N_\mathrm{bis} = 3$. This saturation is likely due to the specific target location, which lies along the line connecting two \acp{BS} in the monostatic configuration and along the direct path between a \ac{Tx}-\ac{Rx} pair in the multistatic setup—limiting the contribution of additional sensing nodes to velocity estimation. 

\emph{Remark:}
This result is presented to emphasize that, while cooperation is always beneficial in principle, it is not necessarily effective in all situations. A framework such as the one proposed in this work is therefore essential for interpreting the behavior of cooperation and for guiding network design decisions.

Fig.~\ref{fig:VEBvsM} presents the \ac{VEB} as a function of the fraction $\rho_t$ of \ac{OFDM} symbols allocated to sensing, comparing cooperative monostatic configurations with $\Nbs = 2, 3, 4$ \acp{BS} and multistatic configurations with $N_\mathrm{bis} = 2, 3$ bistatic pairs. For both architectures, increasing the number of \ac{OFDM} symbols leads to improved velocity estimation accuracy, as indicated by lower \ac{VEB} values. This improvement is due to the fact that a larger number of symbols enables more accurate estimation of the Doppler frequency, and thus of the radial velocity at each \ac{Rx}.

Fig.~\ref{fig:VEBvsNr} illustrates the \ac{VEB} as a function of the number of receive antennas $\Nr$, ranging from $8$ to $100$. As observed for the \ac{PEB}, increasing the number of receive antennas also positively impacts velocity estimation accuracy. This is because a larger array enables more accurate estimation of the target's direction, which is crucial for determining both the magnitude and angle of the target's velocity vector.

\emph{Remark:}
These results underscore the importance of jointly optimizing sensing duration and spatial diversity when designing \ac{ISAC} networks for reliable velocity estimation.

\subsection{\Ac{BS} Selection in Cooperative Monostatic Setup}
From a system-level perspective, the bounds derived in this work enable various forms of performance-driven analysis. For example, in a scenario where $\Nbs$ monostatic \acp{BS} are deployed, these bounds can guide the selection of an optimal subset of \acp{BS} that ensures adequate coverage or maximizes estimation accuracy within a specific region.

Table~\ref{tab:bssel} shows the selection of $4$ \acp{BS} from a cooperative monostatic network consisting of $8$ \acp{BS}, to optimize different estimation metrics at three target locations: $[12, 51]\,$m, $[42, 42]\,$m, and $[70, 20]\,$m. Rather than reporting numerical performance values, each cell in the table uses symbolic markers to indicate whether a given \ac{BS} is part of the optimal subset for a particular metric:
\begin{itemize}
    \item a green dot ({\color{green}\ding{108}}) indicates selection for minimizing the \ac{PEB};
    \item a red square ({\color{red}\ding{110}}) for minimizing the \ac{VEB};
    \item a blue diamond ({\color{blue}\ding{117}}) for minimizing the CRLB of the velocity angle, $\CRB(\angle \mathbf{v})$.
\end{itemize}

The table highlights how different estimation objectives may lead to different \ac{BS} selections. For instance, the \ac{BS} located at $[42, 0]\,$m is consistently selected across almost all metrics and target positions, whereas other \acp{BS} (e.g., those at $[0, 84]\,$m or $[84, 84]\,$m) contribute only to specific metrics at particular locations. When the target is located near the center of the monitored area (around $[42, 42]\,$m), the optimal selection typically includes the four \acp{BS} closest to the center. In contrast, for targets positioned farther from the center, the optimal subset varies more significantly and depends on the specific performance metric being optimized.

\emph{Remark:}
These findings suggest that the optimal subset of \acp{BS} should be selected not only based on geometric proximity but also according to the specific estimation objective, whether it is position accuracy, velocity magnitude, or velocity angle. This underscores the importance of metric-aware and context-dependent \ac{BS} selection strategies in cooperative sensing systems.

\begin{table}[t!]
\centering
\tiny
\renewcommand{\arraystretch}{1.0}
\setlength{\heavyrulewidth}{0.08em}
\setlength{\lightrulewidth}{0.05em}
\setlength{\cmidrulewidth}{0.03em}
\caption{\Ac{BS} selection in cooperative monostatic setup per target position and metric ({\color{green}\ding{108}}: PEB, {\color{red}\ding{110}}: VEB, {\color{blue}\ding{117}}: $\CRB(\angle{\mathbf{v}})$)} 
\label{tab:bssel}
\begin{adjustbox}{width=0.98\columnwidth}
\begin{tabular}{l|ccc}
\toprule
\shortstack{\textbf{Target position [m] $\rightarrow$} \\ \textbf{BS position [m]}  $\downarrow$} &  $[12, 51]$ & $[42, 42]$ & $[70, 20]$ \\
\midrule
 $[42, 0]$      &  {\color{red}\ding{110}}{\color{blue}\ding{117}} 
               & {\color{green}\ding{108}}{\color{red}\ding{110}}{\color{blue}\ding{117}} 
               & {\color{green}\ding{108}}{\color{red}\ding{110}}{\color{blue}\ding{117}} \\
$[0, 42]$      &  {\color{green}\ding{108}}{\color{red}\ding{110}}{\color{blue}\ding{117}} 
               & {\color{green}\ding{108}}{\color{red}\ding{110}}{\color{blue}\ding{117}} 
               & {\color{green}\ding{108}}  \\
$[84, 42]$    &  & {\color{green}\ding{108}}{\color{red}\ding{110}}{\color{blue}\ding{117}} 
               & {\color{green}\ding{108}}{\color{red}\ding{110}}{\color{blue}\ding{117}} \\
$[42, 84]$    &   {\color{green}\ding{108}}          & {\color{green}\ding{108}}{\color{red}\ding{110}}{\color{blue}\ding{117}} 
               &  \\
$[0, 0]$       & {\color{green}\ding{108}}{\color{red}\ding{110}}{\color{blue}\ding{117}} & & {\color{blue}\ding{117}}\\
$[0, 84]$     &   {\color{green}\ding{108}}{\color{red}\ding{110}}{\color{blue}\ding{117}} &  &  \\
$[84, 84]$   &  &  &  {\color{red}\ding{110}} \\
$[84, 0]$     &  &  & {\color{green}\ding{108}}{\color{red}\ding{110}}{\color{blue}\ding{117}} \\
\bottomrule
\end{tabular}
\end{adjustbox}
\end{table}

\subsection{Transmitter Selection in Multistatic Setup}
In a multistatic configuration, beyond selecting the set of participating \acp{BS}, an additional system-level design question concerns the choice of the optimal transmitting \ac{BS}. Given that only one \ac{BS} acts as a \ac{Tx} while the remaining $N_\mathrm{bis}$ nodes operate as \acp{Rx}, this choice can significantly influence the system’s localization and velocity estimation performance.

We consider a scenario composed of $N_\mathrm{bis}+1=8$ \acp{BS}, whose system parameters are reported in Table~\ref{tab:sim_param}. For each candidate \ac{Tx}, we compute the \ac{PEB}, the \ac{VEB}, and the \ac{CRLB} on the velocity angle, assuming all remaining \acp{BS} act as \acp{Rx}. 
The results are summarized in Table~\ref{tab:txsel}, where the optimal \ac{Tx} is highlighted for each metric: \ac{PEB} ({\color{green}\ding{108}}), \ac{VEB} ({\color{red}\ding{110}}), and $\CRB(\angle{\mathbf{v}})$ ({\color{blue}\ding{117}}). The notation follows the same convention as in Table~\ref{tab:bssel}.

From the table, we observe that the optimal \ac{Tx} depends strongly on the target position. For example, at $[70,20]\,$m, different \acp{BS} yield optimal performance depending on the metric, indicating a trade-off between position and velocity estimation accuracy. Additionally, some \acp{BS} (e.g., $[42,0]\,$m and $[84,84]\,$m) are never selected as optimal \acp{Tx}, suggesting their position may be suboptimal for any of the considered bounds and target positions. 

\emph{Remark:}
This type of analysis provides useful design insights for multistatic systems, as it highlights how \ac{Tx} selection can be tailored to specific sensing goals and spatial conditions.

\begin{table}[t!]
\centering
\tiny
\renewcommand{\arraystretch}{1}
\setlength{\heavyrulewidth}{0.08em}
\setlength{\lightrulewidth}{0.05em}
\setlength{\cmidrulewidth}{0.03em}
\caption{Tx selection in multistatic setup per target position and metric ({\color{green}\ding{108}}: PEB, {\color{red}\ding{110}}: VEB, {\color{blue}\ding{117}}: $\CRB(\angle{\mathbf{v}})$)} 
 \label{tab:txsel}
\begin{adjustbox}{width=0.98\columnwidth}
\begin{tabular}{l|@{\hskip 4pt}c@{\hskip 4pt}c@{\hskip 4pt}c@{\hskip 4pt}c@{}}
\toprule
\shortstack{\textbf{Target position [m] $\rightarrow$} \\ \textbf{Tx position [m]}  $\downarrow$} &  $[13, 21]$ & $[23, 64]$ & $[70, 20]$ & $[37, 50]$ \\
\midrule
$[42, 0]$      &  &  & & \\
$[0, 42]$      & {\color{green}\ding{108}}{\color{red}\ding{110}}{\color{blue}\ding{117}} &  &  &  \\
$[84, 42]$     & & & {\color{green}\ding{108}} & \\
$[42, 84]$     &  & {\color{green}\ding{108}} & & {\color{green}\ding{108}}{\color{red}\ding{110}}{\color{blue}\ding{117}} \\
$[0, 0]$       & &  & & \\
$[0, 84]$      &  & {\color{red}\ding{110}}{\color{blue}\ding{117}} & &  \\
$[84, 84]$     & & & & \\
$[84, 0]$      & & & {\color{red}\ding{110}}{\color{blue}\ding{117}} & \\
\bottomrule
\end{tabular}
\end{adjustbox}
\end{table}

\section{Conclusion}\label{sec:conclusions}
In this work, we have derived and analyzed information-theoretic bounds on the performance of target localization and velocity estimation in cooperative networks employing \ac{MIMO}-\ac{OFDM} technology. The proposed framework is general and applies to monostatic, bistatic, and multistatic sensing configurations with arbitrary sensor geometries, while accounting for key system parameters such as bandwidth, power, array size, and observation time.

The resulting closed-form expressions for the \ac{CRLB} provide a rigorous tool to benchmark sensing performance in \ac{ISAC} networks. Numerical evaluations demonstrate that cooperation among \acp{BS} significantly enhances estimation accuracy and extends sensing coverage, even under resource constraints. Moreover, the analysis offers several design insights: for example, the number of receive antennas plays a critical role in localization accuracy; sensor selection based on error bounds can support efficient resource allocation; and configurations that are optimal for position estimation may not be optimal for velocity estimation.

Overall, the proposed framework offers a tractable and theoretically grounded foundation to guide the design and optimization of next-generation wireless systems with distributed sensing capabilities.

\appendices
\section{Derivation of the Fisher Information Matrix} \label{sec:appA}
The \ac{FIM} of the vector of parameters ${\boldsymbol{\Theta}}=(\alpha,\phi, \fd,\tau,\theta_\mathrm{R})$ is calculated from \eqref{eq:LLF} as
\begin{equation}
[\I(\boldsymbol{\Theta})]_{i,j}=-\EX{\frac{\partial^2 \ln f(\boldsymbol{\mathcal Y})}{\partial \theta_i \partial\theta_j}} \quad i,j=1,\dots,5
\end{equation}
which can be rewritten as 
\begin{align}
&[\I
(\boldsymbol{\Theta})]_{i,j} =\mathbb{E} \left\{ \frac{\partial^2}{\partial \theta_i \partial\theta_j} \left[ \sum_{k=0}^{K-1} \sum_{m=0}^{M-1}\frac{1}{\widetilde{\sigma}_\mathrm{N}^2} \left\| \y[k,m] \right.\right.\right.\nonumber\\
& \left.\left.\left. -\alpha e^{\imath \phi} e^{\imath 2\pi m \Ts f_{\mathrm{D}}}e^{-\imath 2\pi k \Df \tau} 
\mathbf{b}(\theta_{\mathrm{R}})
\mathbf{a}^\herm(\theta_{\mathrm{T}})
\mathbf{w}_\mathrm{T}
\right\|^2 \right]\right\}\nonumber\\
& = \sum_{k=0}^{K-1} \sum_{m=0}^{M-1}\mathbb{E} \left\{ \frac{1}{\widetilde{\sigma}_\mathrm{N}^2} 
\frac{\partial^2}{\partial \theta_i \partial\theta_j} \left[ 
\left\| \y[k,m] \right.\right.\right.\nonumber\\
& \left.\left.\left. -\alpha e^{\imath \phi} e^{\imath 2\pi m \Ts f_{\mathrm{D}}}e^{-\imath 2\pi k \Df \tau} 
\mathbf{b}(\theta_{\mathrm{R}})
\mathbf{a}^\herm(\theta_{\mathrm{T}})
\mathbf{w}_\mathrm{T}
\right\|^2 \right]\right\}\nonumber\\
& = \sum_{k=0}^{K-1} \sum_{m=0}^{M-1}\mathbb{E} \left\{ \frac{1}{\widetilde{\sigma}_\mathrm{N}^2} 
\frac{\partial^2}{\partial \theta_i \partial\theta_j}
\left\| \y[k,m] -\beta\h
\right\|^2\right\}
\end{align}
where $\h= \mathbf{b}(\theta_{\mathrm{R}})
\mathbf{a}^\herm(\theta_{\mathrm{T}})
\mathbf{w}_\mathrm{T}=\bv \gamma$ where $\gamma=\mathbf{a}^\herm(\theta_{\mathrm{T}})
\mathbf{w}_\mathrm{T}$ and 
\begin{equation}
\beta=\alpha e^{\imath \phi} e^{\imath 2\pi m \Ts f_{\mathrm{D}}}e^{-\imath 2\pi k \Df \tau}
\end{equation}
with $|\beta|^2=\alpha^2$. The inner differentiation can be conveniently simplified as the sum of three terms
\begin{align}
    \frac{\partial^2}{\partial \theta_i \partial\theta_j}
&\left\| \y[k,m] -\beta\h
\right\|^2\nonumber\\
&=\frac{\partial^2}{\partial \theta_i \partial\theta_j}
\left[|\beta|^2 \hh\h-\beta\yh\h-\beta^* \hh \y
\right]\nonumber\\
&=A_{i,j}-B_{i,j}-C_{i,j}
\end{align}
where
\begin{align}
A_{i,j}&=\frac{\partial^2}{\partial \theta_i \partial\theta_j}
\left[\alpha^2 \hh\h\right]\nonumber\\
B_{i,j}&=\frac{\partial^2}{\partial \theta_i \partial\theta_j}
\left[\beta\yh\h
\right]\nonumber\\
C_{i,j}&=\frac{\partial^2}{\partial \theta_i \partial\theta_j}
\left[\beta^* \hh \y
\right].
\end{align}

\subsection{Amplitude $\alpha$}
It is easy to show that $A_{1,1}=\frac{\partial^2}{\partial\alpha^2}\left[\alpha^2 \hh\h\right]=2\hh\h$, $B_{1,1}=\frac{\partial^2}{\partial\alpha^2}\left[\beta\yh\h\right]=0$ and $C_{1,1}=\frac{\partial^2}{\partial\alpha^2}\left[\beta^*\hh\y\right]=0$, hence
\begin{align}
\mathcal{I}_{\alpha,\alpha}=\sum_{k=0}^{K-1} \sum_{m=0}^{M-1}\mathbb{E} \left\{ \frac{1}{\widetilde{\sigma}_\mathrm{N}^2}
2\hh\h\right\}=2KM\frac{\|\h\|^2}{\widetilde{\sigma}_\mathrm{N}^2}.
\end{align}

As far as the entry $(1,2)$, we can derive $A_{1,2}=\frac{\partial^2}{\partial\alpha \partial \phi}\left[\alpha^2 \hh\h\right]=0$, $B_{1,2}=\frac{\partial^2}{\partial\alpha \partial \phi}\left[\beta\yh\h\right]=\imath \beta\yh\h/\alpha$ and $C_{1,2}=\frac{\partial^2}{\partial\alpha \partial \phi}\left[\beta^*\hh\y\right]=-\imath\beta^*\hh\y/\alpha$, hence
\begin{align}
\mathcal{I}_{\alpha,\phi}&=\sum_{k=0}^{K-1} \sum_{m=0}^{M-1}\mathbb{E} \left\{ \frac{1}{\widetilde{\sigma}_\mathrm{N}^2}(
-\imath \beta\yh\h/\alpha+\imath\beta^*\hh\y/\alpha)\right\}\nonumber\\
&=\frac{KM}{\widetilde{\sigma}_\mathrm{N}^2}
(
-\imath \beta\EX{\yh}\h/\alpha+\imath\beta^*\hh\EX{\y}/\alpha)
\nonumber\\
&=\frac{KM}{\widetilde{\sigma}_\mathrm{N}^2}\|\h\|^2
(
-\imath |\beta|^2/\alpha+\imath|\beta|^2/\alpha)=0.
\end{align}

As far as the entry $(1,3)$, we can derive $A_{1,3}=\frac{\partial^2}{\partial\alpha \partial \fd}\left[\alpha^2 \hh\h\right]=0$, $B_{1,3}=\frac{\partial^2}{\partial\alpha \partial \fd}\left[\beta\yh\h\right]=\imath 2 \pi m \Ts \beta\yh\h/\alpha$ and $C_{1,3}=\frac{\partial^2}{\partial\alpha \partial \fd}\left[\beta^*\hh\y\right]=-\imath2 \pi m \Ts\beta^*\hh\y/\alpha$, hence
\begin{align}
\mathcal{I}_{\alpha,\fd}&=\sum_{k=0}^{K-1} \sum_{m=0}^{M-1}\mathbb{E} \left\{\frac{1}{\widetilde{\sigma}_\mathrm{N}^2}2\pi m \Ts(
-\imath \beta\yh\h/\alpha\right.\nonumber\\
&\left.+\imath\beta^*\hh\y/\alpha)\right\}\nonumber\\
&=\frac{K\sum_{m=0}^{M-1} m}{\widetilde{\sigma}_\mathrm{N}^2}2 \pi \Ts
(
-\imath \beta\EX{\yh}\h/\alpha+\imath\beta^*\hh\EX{\y}/\alpha)
\nonumber\\
&=\frac{KM(M-1)}{2\widetilde{\sigma}_\mathrm{N}^2}\|\h\|^2
2 \pi \Ts(
-\imath |\beta|^2/\alpha+\imath|\beta|^2/\alpha)=0.
\end{align}

As far as the entry $(1,4)$, the derivation of $\mathcal{I}_{\alpha,\tau}$ follows the same structure (in $\beta$, $\fd$ and $\tau$ are almost interchangeable), leading to $\mathcal{I}_{\alpha,\tau}=0$. 

As far as the entry $(1,5)$, we can derive $A_{1,5}=\frac{\partial^2}{\partial\alpha \partial \thetar}\left[\alpha^2 \hh\h\right]=
\frac{\partial}{\partial \thetar}
\left[2\alpha |\gamma|^2\bh(\thetar)\bv(\thetar)\right]=\frac{\partial}{\partial \thetar}
\left[2\alpha |\gamma|^2\Nr\right]=0$, 
$B_{1,5}=\frac{\partial^2}{\partial\alpha \partial \thetar}\left[\beta\yh\h\right]=\frac{\beta\gamma}{\alpha}\yh\bp(\thetar)$ 
and 
$C_{1,5}=\frac{\partial^2}{\partial\alpha \partial \thetar}\left[\beta^*\hh\y\right]=\frac{\beta^*\gamma^*}{\alpha}\bph(\thetar)\y$, hence
\begin{align}
\mathcal{I}_{\alpha,\thetar}&=\sum_{k=0}^{K-1} \sum_{m=0}^{M-1}\mathbb{E} \left\{ \frac{1}{\widetilde{\sigma}_\mathrm{N}^2}\left(
-\frac{\beta\gamma}{\alpha}\yh\bp(\thetar)\right.\right.\nonumber\\
&\left.\left.-\frac{\beta^*\gamma^*}{\alpha}\bph(\thetar)\y\right)
\right\}\nonumber\\
&=\frac{KM}{\widetilde{\sigma}_\mathrm{N}^2}
\left(-\frac{|\beta|^2\gamma}{\alpha}\hh\bp(\thetar)
-\frac{|\beta|^2\gamma^*}{\alpha}\bph(\thetar)\h
\right)
\nonumber\\
&=-\frac{KM}{\widetilde{\sigma}_\mathrm{N}^2}\alpha
\left(\gamma\hh\bp(\thetar)
+\gamma^*\bph(\thetar)\h
\right)\nonumber\\
&=-\frac{2KM}{\widetilde{\sigma}_\mathrm{N}^2}\alpha
\Re\left\{\gamma\hh\bp(\thetar)\right\}=0
\end{align}
where the last equality is due to the orthogonal property \eqref{eq:orthogprop}, i.e., $\hh\bp(\thetar)=\gamma^*\bh(\thetar)\bp(\thetar)=0$.

\subsection{Target and Channel Phase $\phi$}
It is easy to show that $A_{2,2}=\frac{\partial^2}{\partial\phi^2}\left[\alpha^2 \hh\h\right]=0$, $B_{2,2}=\frac{\partial^2}{\partial\phi^2}\left[\beta\yh\h\right]=-\beta\yh\h$ and $C_{2,2}=\frac{\partial^2}{\partial\phi^2}\left[\beta^*\hh\y\right]=-\beta^*\hh\y$, hence
\begin{align}
\mathcal{I}_{\phi,\phi}&=\sum_{k=0}^{K-1} \sum_{m=0}^{M-1}\mathbb{E} \left\{ \frac{1}{\widetilde{\sigma}_\mathrm{N}^2}
(\beta\yh\h+\beta^*\hh\y))\right\}=\nonumber\\
&=2KM\frac{\alpha^2\|\h\|^2}{\widetilde{\sigma}_\mathrm{N}^2}.
\end{align}

As far as the entry $(2,3)$, we can derive $A_{2,3}=\frac{\partial^2}{\partial\phi \partial \fd}\left[\alpha^2 \hh\h\right]=0$, $B_{2,3}=\frac{\partial^2}{\partial\phi \partial \fd}\left[\beta\yh\h\right]=-2\pi m \Ts \beta \yh\h$ and $C_{2,3}=\frac{\partial^2}{\partial\phi \partial \fd}\left[\beta^*\hh\y\right]=-2\pi m \Ts \beta^*\hh\y$, hence
\begin{align}
\mathcal{I}_{\phi,\fd}&=\sum_{k=0}^{K-1} \sum_{m=0}^{M-1}\mathbb{E} \left\{ \frac{1}{\widetilde{\sigma}_\mathrm{N}^2}(
-\imath \beta\yh\h/\alpha+\imath\beta^*\hh\y/\alpha)\right\}\nonumber\\
&=\frac{2\pi K\Ts}{\widetilde{\sigma}_\mathrm{N}^2}
\left(\sum_{m=0}^{M-1}m\right)(\beta\EX{\yh}\h+\beta^*\hh\EX{\y})\nonumber\\
&=2\pi K \Ts M(M-1)
\frac{\alpha^2 \|\h\|^2}{\widetilde{\sigma}_\mathrm{N}^2}
\end{align}

As far as the entry $(2,4)$, we can derive $A_{2,4}=\frac{\partial^2}{\partial\phi \partial \tau}\left[\alpha^2 \hh\h\right]=0$, $B_{2,4}=\frac{\partial^2}{\partial\phi \partial \tau}\left[\beta\yh\h\right]=2\pi k \Df \beta \yh\h$ and $C_{2,4}=\frac{\partial^2}{\partial\phi \partial \tau}\left[\beta^*\hh\y\right]=2\pi k \Df \beta^*\hh\y$, hence
\begin{align}
\mathcal{I}_{\phi,\tau}&=-\sum_{k=0}^{K-1} \sum_{m=0}^{M-1}\mathbb{E} \left\{ \frac{1}{\widetilde{\sigma}_\mathrm{N}^2}2\pi k \Df(\beta\yh\h+\beta^*\hh\y)\right\}\nonumber\\
&=-\frac{2\pi M\Df}{\widetilde{\sigma}_\mathrm{N}^2}
\left(\sum_{k=0}^{K-1}k\right)(\beta\EX{\yh}\h+\beta^*\hh\EX{\y})\nonumber\\
&=-2\pi M \Df K(K-1)
\frac{\alpha^2 \|\h\|^2}{\widetilde{\sigma}_\mathrm{N}^2}
\end{align}

As far as the entry $(2,5)$, we can derive $A_{2,5}=\frac{\partial^2}{\partial\phi \partial \thetar}\left[\alpha^2 \hh\h\right]=0$, 
$B_{2,5}=\frac{\partial^2}{\partial\phi \partial \thetar}\left[\beta\yh\h\right]=\imath \beta 
\frac{\partial}{\partial \thetar}\left[
\yh \bv(\thetar)\gamma\right]=
\imath \beta \gamma \yh \bp(\thetar)$ 
and 
$C_{2,5}=\frac{\partial^2}{\partial\phi \partial \thetar}\left[\beta^*\hh\y\right]=-\imath \beta^* 
\frac{\partial}{\partial \thetar}\left[
\bh(\thetar)\gamma^*\y\right]=
-\imath \beta^* \gamma^* \bph(\thetar)\y$, hence
\begin{align}
\mathcal{I}_{\phi,\thetar}&=\sum_{k=0}^{K-1} \sum_{m=0}^{M-1}\mathbb{E} \left\{ \frac{1}{\widetilde{\sigma}_\mathrm{N}^2}\left(
-\imath\beta\gamma\yh\bp(\thetar)\right.\right.\nonumber\\
&\left.\left.+\imath\beta^*\gamma^*\bph(\thetar)\y\right)
\right\}\nonumber\\
&=\frac{KM}{\widetilde{\sigma}_\mathrm{N}^2}
\left(-\imath |\beta|^2\gamma\hh\bp(\thetar)
+\imath |\beta|^2\gamma^*\bph(\thetar)\h
\right)
\nonumber\\
&=\frac{2KM}{\widetilde{\sigma}_\mathrm{N}^2}\alpha^2
\Im\left\{\gamma\hh\bp(\thetar)\right\}=0
\end{align}
where the last equality is due to the orthogonal property \eqref{eq:orthogprop}, i.e., $\hh\bp(\thetar)=\gamma^*\bh(\thetar)\bp(\thetar)=0$.

\subsection{Doppler Frequency $\fd$}
It is easy to show that $A_{3,3}=\frac{\partial^2}{\partial \fd^2}\left[\alpha^2 \hh\h\right]=0$, $B_{3,3}=\frac{\partial^2}{\partial\fd^2}\left[\beta\yh\h\right]=-4\pi^2 m^2 \Ts^2\beta\yh\h$ and $C_{3,3}=\frac{\partial^2}{\partial\fd^2}\left[\beta^*\hh\y\right]=-4\pi^2 m^2 \Ts^2 \beta^*\hh\y$, hence
\begin{align}
\mathcal{I}_{\fd,\fd}&=\sum_{k=0}^{K-1} \sum_{m=0}^{M-1} 4 \pi ^2 \Ts^2 m^2 \mathbb{E} \left\{ \frac{1}{\widetilde{\sigma}_\mathrm{N}^2}
(\beta\yh\h \right.\nonumber\\
&\left.+\beta^*\hh\y))\right\}=\nonumber\\
&=\frac{4\pi^2 K \Ts^2}{\sigma^2_\mathrm{N}} \left(\sum_{m=0}^{M-1}m^2\right)(\beta\EX{\yh}\h+\beta^*\hh\EX{\y})\nonumber\\
&=8\pi^2 K \Ts^2 \frac{(2M-1)(M-1)M}{6}
\frac{|\beta|^2\|\h\|^2}{\widetilde{\sigma}_\mathrm{N}^2}\nonumber\\
&=4\pi^2 K \Ts^2 \frac{(2M-1)(M-1)M}{3}
\frac{\alpha^2\|\h\|^2}{\widetilde{\sigma}_\mathrm{N}^2}.
\end{align}

As far as the entry $(3,4)$, we can derive $A_{3,4}=\frac{\partial^2}{\partial\fd \partial \tau}\left[\alpha^2 \hh\h\right]=0$, $B_{3,4}=\frac{\partial^2}{\partial\fd \partial \tau}\left[\beta\yh\h\right]=4\pi^2 m k \Ts \Df \beta \yh\h$ and $C_{3,4}=\frac{\partial^2}{\partial\fd \partial \tau}\left[\beta^*\hh\y\right]=4\pi^2 m k \Ts \Df \beta^*\hh\y$, hence
\begin{align}
\mathcal{I}_{\fd,\tau}&=\sum_{k=0}^{K-1} \sum_{m=0}^{M-1}4 \pi^2 m k \Ts \Df \mathbb{E} \left\{ \frac{1}{\widetilde{\sigma}_\mathrm{N}^2}(
-\beta\yh\h\right.\nonumber\\
&\left.-\beta^*\hh\y)\right\}\nonumber\\
&=-\frac{4\pi^2 \Ts \Df}{\widetilde{\sigma}_\mathrm{N}^2}
\left(\sum_{k=0}^{K-1}k\right)\left(\sum_{m=0}^{M-1}m\right)(\beta\EX{\yh}\h\nonumber\\
&+\beta^*\hh\EX{\y})\nonumber\\
&=-2\pi^2 \Ts \Df K(K-1) M(M-1)
\frac{\alpha^2 \|\h\|^2}{\widetilde{\sigma}_\mathrm{N}^2}.
\end{align}

As far as the entry $(3,5)$, we can derive $A_{3,5}=\frac{\partial^2}{\partial\fd \partial \thetar}\left[\alpha^2 \hh\h\right]=0$, $B_{3,5}=\frac{\partial^2}{\partial\fd \partial \thetar}\left[\beta\yh\h\right]=\frac{\partial}{\partial \thetar}[\imath 2 \pi m \Ts \beta \yh h]=\imath 2 \pi m \Ts \beta \gamma \yh \bp(\thetar)$ and $C_{3,5}=\frac{\partial^2}{\partial\fd \partial \thetar}\left[\beta^*\hh\y\right]=\frac{\partial}{\partial \thetar}[-\imath 2 \pi m \Ts \beta^* \hh \y]=-\imath 2 \pi m \Ts \beta^* \gamma^* \bph(\thetar)\y$, hence
\begin{align}
\mathcal{I}_{\fd,\thetar}&=-\sum_{k=0}^{K-1} \sum_{m=0}^{M-1} \imath 2 \pi m \Ts \mathbb{E}\left\{ \frac{1}{\widetilde{\sigma}_\mathrm{N}^2}(\beta\gamma\yh\bp(\thetar)\right.\nonumber\\
&\left.-\beta^*\gamma^*\bph(\thetar)\y)\right\}\nonumber\\
&=-\imath\frac{2\pi K\Ts}{\widetilde{\sigma}_\mathrm{N}^2}
\left(\sum_{m=0}^{M-1}m\right)(\beta\gamma\EX{\yh}\bp(\thetar)\nonumber\\
&-\beta^*\gamma^*\bph(\thetar)\EX{\y})\nonumber\\
&=\frac{2\pi K \Ts\alpha^2}{\widetilde{\sigma}_\mathrm{N}^2}M(M-1)\Im\{\gamma\hh\bp(\thetar)\}=0
\end{align}
where the last equality is due to the orthogonal property \eqref{eq:orthogprop}, i.e., $\hh\bp(\thetar)=\gamma^*\bh(\thetar)\bp(\thetar)=0$.

\subsection{Time Delay $\tau$}
It is easy to show that $A_{4,4}=\frac{\partial^2}{\partial \tau^2}\left[\alpha^2 \hh\h\right]=0$, $B_{4,4}=\frac{\partial^2}{\partial\tau^2}\left[\beta\yh\h\right]=-4\pi^2 k^2 \Df^2\beta\yh\h$ and $C_{4,4}=\frac{\partial^2}{\partial\tau^2}\left[\beta^*\hh\y\right]=-4\pi^2 k^2 \Df^2 \beta^*\hh\y$, hence
\begin{align}
\mathcal{I}_{\tau,\tau}&=\sum_{k=0}^{K-1} \sum_{m=0}^{M-1} 4 \pi ^2 \Df^2 k^2 \mathbb{E} \left\{ \frac{1}{\widetilde{\sigma}_\mathrm{N}^2}
(\beta\yh\h \right.\nonumber\\
&\left.+\beta^*\hh\y))\right\}=\nonumber\\
&=\frac{4\pi^2 M \Df^2}{\widetilde{\sigma}^2_\mathrm{N}} \left(\sum_{k=0}^{K-1}k^2\right)(\beta\EX{\yh}\h+\beta^*\hh\EX{\y})\nonumber\\
&=8\pi^2 M \Df^2 \frac{(2K-1)(K-1)K}{6}
\frac{|\beta|^2\|\h\|^2}{\widetilde{\sigma}_\mathrm{N}^2}\nonumber\\
&=4\pi^2 M \Df^2 \frac{(2K-1)(K-1)K}{3}
\frac{\alpha^2\|\h\|^2}{\widetilde{\sigma}_\mathrm{N}^2}.
\end{align}

As far as the entry $(4,5)$, the approach is very similar to that followed for the entry $(3,5)$ because the role of $\fd$ and $\tau$ is interchangeable, hence, after some lengthy calculations: 
%
$\mathcal{I}_{\tau,\thetar}=0.$

\subsection{AoA $\thetar$}
It is easy to show that $A_{5,5}=\frac{\partial^2}{\partial \thetar^2}\left[\alpha^2 \hh\h\right]=\alpha^2|\gamma|^2 \frac{\partial^2}{\partial \thetar^2}\left[\bh(\thetar)\bv(\thetar)\right]=0$ because of \eqref{eq:steerprop}, $B_{5,5}=\frac{\partial^2}{\partial\thetar^2}\left[\beta\yh\h\right]=\beta\gamma\yh\bpp(\thetar)$ and $C_{5,5}=\frac{\partial^2}{\partial\thetar^2}\left[\beta^*\hh\y\right]=\beta^*\gamma^*\bpph(\thetar)\y$, hence
\begin{align}
\mathcal{I}_{\thetar,\thetar}&=\sum_{k=0}^{K-1} \sum_{m=0}^{M-1} \mathbb{E} \left\{ \frac{1}{\widetilde{\sigma}_\mathrm{N}^2}
(-\beta\gamma\yh\bpp(\thetar) \right.\nonumber\\
&\left.-\beta^*\gamma^*\bpph(\thetar)\y)\right\}=\nonumber\\
&=\frac{KM}{\widetilde{\sigma}^2_\mathrm{N}} (-\beta\gamma\EX{\yh}\bpp(\thetar)-\beta^*\gamma^*\bpph(\thetar)\EX{\y})\nonumber\\
&=\frac{KM\alpha^2}{\widetilde{\sigma}^2_\mathrm{N}} (-\gamma\hh\bpp(\thetar)-\gamma^*\bpph(\thetar)\h)\nonumber\\
&=-\frac{2KM\alpha^2}{\widetilde{\sigma}^2_\mathrm{N}} \Re\{\gamma\hh\bpp(\thetar)\}\nonumber\\
&=\frac{\pi^2KM\alpha^2|\gamma|^2}{6\widetilde{\sigma}^2_\mathrm{N}} (\Nr^2-1)\Nr \cos^2(\thetar)
\end{align}
where the last equality accounts for the property \eqref{eq:bppprop}.

\balance
\bibliographystyle{./bibliography/IEEEtran}
\bibliography{./bibliography/IEEEabrv,./bibliography/Bibliography}

\end{document}